\documentclass[aps,pre,twocolumn,superscriptaddress,showpacs]{revtex4}
\usepackage{graphicx}
\usepackage{chemarr}
\usepackage{amsthm}
\begin{document}

\title{Quantifying Stochastic Effects in Biochemical Reaction Networks using Partitioned Leaping}

\author{Leonard A. Harris}
\altaffiliation[Current address: ]{Department of Computational Biology, University of Pittsburgh School of Medicine, Pittsburgh, PA 15260, USA.}%
\email{lharris@pitt.edu}%
\affiliation{School of Chemical and Biomolecular Engineering, Cornell University, Ithaca, NY 14853, USA}

\author{Aaron M. Piccirilli}
\affiliation{School of Chemical and Biomolecular Engineering, Cornell University, Ithaca, NY 14853, USA}%
\affiliation{Department of Computer Science, Cornell University, Ithaca, NY 14853, USA}

\author{Emily R. Majusiak}
\affiliation{School of Chemical and Biomolecular Engineering, Cornell University, Ithaca, NY 14853, USA}

\author{Paulette Clancy}%
\email{pqc1@cornell.edu}%
\affiliation{School of Chemical and Biomolecular Engineering, Cornell University, Ithaca, NY 14853, USA}

\date{\today}

\begin{abstract}
``Leaping" methods show great promise for significantly accelerating stochastic simulations of complex biochemical reaction networks. However, few practical
applications of leaping have appeared in the literature to date.  Here, we address this issue using the ``partitioned leaping algorithm" (PLA) [L.A.\ Harris
and P.~Clancy, J.~Chem.\ Phys.\ \textbf{125}, 144107 (2006)], a recently-introduced multiscale leaping approach. We use the PLA to investigate stochastic
effects in two model biochemical reaction networks. The networks that we consider are simple enough so as to be accessible to our intuition but sufficiently
complex so as to be generally representative of real biological systems. We demonstrate how the PLA allows us to quantify subtle effects of stochasticity in
these systems that would be difficult to ascertain otherwise as well as not-so-subtle behaviors that would strain commonly-used ``exact" stochastic methods.
We also illustrate bottlenecks that can hinder the approach and exemplify and discuss possible strategies for overcoming them. Overall, our aim is to aid and
motivate future applications of leaping by providing stark illustrations of the benefits of the method while at the same time elucidating obstacles that are
often encountered in practice.
\end{abstract}


\maketitle

\section{Introduction} \label{sec:Intro}

Biological systems are inherently noisy, or \emph{stochastic\/}. A primary source of this noise is the random nature of molecular interactions that
predominates when molecular copy numbers are low, so-called ``intrinsic" noise \citep{McAdams99, Rao02, Raser05, Kaern05, Vent06, Samo06:Dev, Samo06:STKE,
Mahesh07}. Intrinsic noise has been implicated as the source of non-genetic variability in clonal cell populations \citep{Spud76, Elow02, Fedo02} and can
profoundly affect the dynamical behavior of a biological system, both to its benefit as well as its detriment \citep{Rao02}. For example, stochastic effects in
gene expression have been shown to speed the response of yeast cells to a challenge, allowing them to survive in conditions where they otherwise could not
\citep{Blake06}. Conversely, noise can interfere with the workings of circadian clocks \citep{Elow00, Barkai00}, systems whose reliability is essential for
survival. As such, Nature has developed regulatory mechanisms to attenuate the noise \citep{Vilar02, Gonze02:PNAS}. All in all, accounting for the effects of
stochasticity is \emph{essential\/} for gaining a clear understanding of the design principles underlying many biological processes.

Nevertheless, biological-systems modeling today continues to be performed primarily at the continuous-deterministic scale, usually in the form of ordinary
differential equations (ODEs). These formalisms explicitly \emph{ignore\/} stochastic fluctuations \citep{Aldr06, Tomlin07}.  However, the practice
understandably persists because the methods are well established and numerous easy-to-use algorithms are available at little to no cost. Moreover, established
stochastic methods, such as Gillespie's stochastic simulation algorithm (SSA) \citep{Gillesp76, Gillesp77, Gillesp07}, are in many cases simply too
computationally intensive to apply to realistic models of biological networks \citep{Endy01, Gillesp01}.

There is great interest, therefore, in developing accelerated stochastic simulation methods that can accurately capture noise effects but at significantly
reduced computational cost relative to standard approaches.  Ultimately, it is hoped that these methods will supplant ODEs as the default method of choice in
computational systems biology.  Approaches that have been proposed in this regard include (but are not limited to): (i) modification and optimization of
Gillespie's original SSA \citep{Gibson00, Resat01, Cao04:ODM, McColl06}, (ii) ``leaping" methods which ignore the exact moments at which reaction firings occur
\citep{Gillesp01, Gillesp03, Rath03, Cao04:Stability, Tian04, Chatt05, Cao05:negPop, Cao06:newStep, Wagner06, Auger06, Cai07, Petti07, Peng07, Cao07, Rath07,
Ander08, Xu08, Leier08, Harris06}, and (iii) ``hybrid" methods which couple different simulation techniques (e.g., the SSA and ODEs) into a single, overarching
algorithmic framework \citep{Hasel02, Kiehl04, Taka04, Vasu04, Burr04, Puch04, Salis05:hybrid, Griff06, Wylie06}. Of these, leaping methods are particularly
popular, presumably because of the sound theoretical foundation on which they stand \citep{Gillesp00, Gillesp01}.

Despite their popularity, however, very few practical applications of leaping have appeared in the literature to date \citep{Petti07}. This is a curious fact
that has yet to be fully explored or explained. Moreover, those applications that have appeared (e.g., \citep{Chatt05:BioInfo, Perc07, Handel07}) are generally
brief in their presentation of the algorithms used and do not report much by way of the capabilities and limitations of the method. Thus, it is difficult, if
not impossible, to project the potential utility of the approach onto other and more complex biological networks.

Here, we address this issue by using the ``partitioned leaping algorithm" (PLA) \citep{Harris06}, a recently-introduced extension and variant of the
$\tau$\/-leaping method of Gillespie and co-workers \citep{Gillesp01, Gillesp03, Cao05:negPop, Cao06:newStep}, to systematically investigate the effects of
stochasticity in two model biochemical reaction networks.  The systems that we consider are intuitively simple yet they contain attributes that are ubiquitous
to complex biological networks, such as enzyme catalysis and feedback control. We perform a detailed and in-depth investigation using leaping with the aim of
illuminating both the capabilities and limitations of the method and, hence, aiding and motivating future applications of the approach. The PLA operates on the
same basic principles that underlie all leaping algorithms and its performance with respect to these systems can thus be seen as generally reflective of the
entire class of method. That being said, there are certain aspects of the PLA, which we will expound upon below, that make it particularly appealing from a
practical point of view.

We begin in Sec.~\ref{sec:Networks} by introducing the biochemical reaction networks that we investigate in this article.  We then briefly describe in
Sec.~\ref{sec:Methods} the PLA, the time series analysis method and the statistical tests employed in this work. Detailed results for the two networks are
subsequently presented in Sec.~\ref{sec:Results}.  We conclude in Sec.~\ref{sec:Discuss} with a discussion of the implications of these results, a possible
explanation for why practical applications of leaping are so scarce in the literature and the future outlook for leaping methods in computational systems
biology.

\section{The Networks} \label{sec:Networks}

The systems that we investigate are a core model for calcium oscillations in hepatocytes introduced by \citet{Kumm00} and the three-gene ``repressilator" of
\citet{Elow00}.  These systems are relatively simple, yet they are not ``toy" problems in the sense that they contain non-trivial features that are ubiquitous
to biochemical systems, such as enzyme catalysis and feedback control.  Moreover, both systems emit large-amplitude oscillations which give rise to the kinds
of wide disparities in species populations that leaping algorithms are specifically designed to cope with \citep{Gillesp01, Harris06}.  All in all, these
systems provide an ideal testbed for investigating the practical utility of leaping methods in computational systems biology.

Our investigation entails using the PLA to probe behavioral changes that arise in these systems due to changes in various system properties. Specifically, we
investigate the transition from stochastic to deterministic behavior that accompanies increases in the system volume (i.e., population levels) in the
calcium-oscillations model and increases in the gene-protein binding and unbinding rate constants in the repressilator. The salient feature of our
investigation is that we are able to ascertain, in a systematic way, the performance characteristics of the leaping algorithm over a wide spectrum of
conditions. We thus identify cases where leaping proves particularly beneficial, where it ``bogs down," and various points in between.  Further details of the
networks are provided in the subsections below.

\subsection{Calcium Oscillations} \label{sec:Networks::CaOsc}

Intracellular calcium is an important second messenger for the functioning of many cell types, both in plants and in animals.  It is involved in a multitude of
functions during the lifetime of a cell, including fertilization, development and death \citep{Berr98}.  The dynamics of intracellular calcium are not smooth
and continuous, however.  Rather, they are driven by small numbers of receptors and ion channels that can give rise to highly stochastic behavior. Indeed,
experiments have shown that calcium waves are triggered by elementary stochastic events known as ``blips" and ``puffs" \citep{Falcke04}. Incorporating
stochasticity into models of calcium oscillations is thus of high interest.

Many theoretical models have been proposed to describe the oscillatory dynamics of intracellular calcium \citep{Schus02, Falcke04}. \citet{Kumm00} proposed a
model for calcium oscillations in hepatocytes (liver cells) that displays a rich variety of behaviors.  The model features self-enhanced activation of the
$G_\alpha$\/ subunit of the receptor complex and is able to capture many aspects of experimentally-observed behavior that eluded previous models. The authors
also presented a simplified version of the model that displays the same basic behaviors as the full model, thus emphasizing the ``core" mechanisms driving the
oscillations \citep{Kumm00}.

In Table~\ref{table:CaRxns}, we show the \citet{Kumm00} core model for calcium oscillations in hepatocytes.  The model consists of eight reactions involving
three species: the activated form of phospholipase~C (PLC$^\ast$), the $\alpha$ subunit of the receptor-bound $G$ protein ($G_\alpha$\/) and cytosolic calcium
ions (Ca).  Note that the model is in a reduced form, with degradation processes described in terms of Michaelis-Menten kinetics.  Reaction~2, which is the
prime feature of this model, describes the agonist-initiated [e.g., adenosine triphosphate (ATP)] autocatalytic activation of the $G_\alpha$\/ subunit. The
parameter $k_2$\/ thus amounts to the product of the second-order association constant and the agonist concentration and is a primary determinant of the system
behavior. \citeauthor{Kumm00} showed that with increasing $k_2$\/ the system behavior transitions from simple Ca$^{2+}$ spiking oscillations, to complex
oscillations, to chaotic behavior and, finally, to an elevated steady state \citep{Kumm00, Kumm05}.

\begin{table*}[t]
\centering%
\caption{\citet{Kumm00} core model for calcium oscillations in hepatocytes.  `$\emptyset$' represents a source or a sink and $k_2\!=\!2.85$~s$^{-1}$ puts the
system into the ``periodic-bursting" regime (see text). $G_\alpha$\/ represents the activated $\alpha$\/-subunit of the intracellular receptor-bound G-protein,
$\mathrm{PLC}^\ast$ the activated form of phospholipase~C, and $\mathrm{Ca}$ cytosolic calcium ions.  Note that to perform stochastic simulations all
parameters must be devoid of molar units (M). Parameters with molar units are thus multiplied by $N_A\Omega$ (Avogadro's number~$\times$ system volume) prior
to runtime.}
\begin{tabular}{ll@{\hspace{20pt}}l@{\hspace{20pt}}l} \hline\hline
    \multicolumn{2}{l}{Reaction} & Rate Expression & Parameter Value(s) \\ \hline
    1.\, & $\emptyset \rightarrow G_\alpha$                             & $k_1$                                                 & $k_1   =0.212$ M~s$^{-1}$ \\
    2.   & $G_\alpha \rightarrow 2G_\alpha$                             & $k_2[G_\alpha]$                                       & $k_2   =2.85 $ s$^{-1}$ \\
    3.   & $G_\alpha + \mathrm{PLC}^\ast \rightarrow \mathrm{PLC}^\ast$ & $k_3[G_\alpha][\mathrm{PLC}^\ast]/(K_4+[G_\alpha])$   & $k_3   =1.52 $ s$^{-1}$,   $K_4   =0.19$~M \\
    4.   & $G_\alpha + \mathrm{Ca} \rightarrow \mathrm{Ca}$             & $k_5[G_\alpha][\mathrm{Ca}]/(K_6+[G_\alpha])$         & $k_5   =4.88 $ s$^{-1}$,   $K_6   =1.18$~M \\
    5.   & $G_\alpha \rightarrow G_\alpha + \mathrm{PLC}^\ast$          & $k_7[G_\alpha]$                                       & $k_7   =1.24 $ s$^{-1}$ \\
    6.   & $\mathrm{PLC}^\ast \rightarrow \emptyset$                    & $k_8[\mathrm{PLC}^\ast]/(K_9+[\mathrm{PLC}^\ast])$    & $k_8   =32.24$ M~s$^{-1}$, $K_9   =29.09$~M \\
    7.   & $G_\alpha \rightarrow G_\alpha + \mathrm{Ca}$                & $k_{10}[G_\alpha]$                                    & $k_{10}=13.58$ s$^{-1}$ \\
    8.   & $\mathrm{Ca} \rightarrow \emptyset$                          & $k_{11}[\mathrm{Ca}]/(K_{12}+[\mathrm{Ca}])$          & $k_{11}=153.0$ M~s$^{-1}$, $K_{12}=0.16$~M \\
    \hline\hline
    \multicolumn{4}{l}{\footnotesize{Initial conditions: $[G_\alpha]=[\mathrm{PLC}^\ast]=[\mathrm{Ca}]=0.01$~M}} \\
\end{tabular}
\label{table:CaRxns}
\end{table*}

We also see in Table~\ref{table:CaRxns} that the model contains various feedback loops which drive the oscillatory behavior of the network. Specifically,
$\mathrm{PLC}^\ast$ and $\mathrm{Ca}$ are created autocatalytically in reactions 5 and 7, respectively, through the action of $G_\alpha$\/. In reactions 3 and
4, however, $G_\alpha$\/ is degraded enzymatically by the actions of $\mathrm{PLC}^\ast$ and $\mathrm{Ca}$, respectively.  Thus, in the correct parameter
range, increased levels of $G_\alpha$\/ lead to increased levels of $\mathrm{PLC}^\ast$ and $\mathrm{Ca}$ which, in turn, lead to increased degradation of
$G_\alpha$\/, which leads to decreased levels of $\mathrm{PLC}^\ast$ and $\mathrm{Ca}$, and so on and so forth.

In Ref.~\citep{Kumm05}, \citeauthor{Kumm05}\ compared the deterministic behavior of this model to results of stochastic simulations performed using the SSA.
The goal was to determine points of transition to determinism for various dynamical regimes of the model (e.g., ``periodic spiking," ``periodic bursting,"
``chaos") and to provide general insight as to when a deterministic treatment is applicable and when a stochastic approach is necessary. SSA simulations were
performed for various system sizes (with fixed concentrations) and the point of transition to determinism was estimated via \emph{visual\/} comparison of
stochastic and deterministic time courses. Visual inspection was necessary because of the high computational expense of the SSA \citep{Kumm05}.

Here, we extend the analysis of \citeauthor{Kumm05}\ for the ``periodic-bursting" regime, a main focus of Ref.~\citep{Kumm05}. The regime is characterized by
complex Ca$^{2+}$ oscillations comprised of three-peak complexes (see below), behavior that is reminiscent of that seen experimentally in hepatocytes
stimulated by ATP \citep{Kumm00, Kumm05, Dixon90}. Using the PLA and the peak-analysis tool described in Sec.~\ref{sec:Methods}, we collect large amounts of
peak amplitude and peak-to-peak distance data at various system volumes and \emph{quantify\/} the relationship between stochasticity and system size, something
that was not feasible in Ref.~\citep{Kumm05} because of the limitations of the SSA. This allows us to pinpoint, from a statistical perspective, the points of
transition to determinism.  As we shall see, these differ, to some extent, from those reported in \citep{Kumm05}.

\subsection{Repressilator} \label{sec:Networks::Repress}

Synthetic biology is a relatively new and rapidly growing scientific field \citep{Hasty01, Hasty02, Sprin05, Benn05}.  In analogy with electrical circuit
design, synthetic biologists attempt to use their knowledge of fundamental biological principles to design and construct artificial biological ``circuits" that
confer novel function unto their host. In this way, one can isolate and control specific aspects of a biological process and circumvent the immense complexity
of natural biological systems, providing a means by which current theoretical understanding can be tested and scrutinized. Moreover, the long-term goal is to
develop protocols for logical control. One can envision a time when microorganisms are ``programmed" at the genetic level to carry out important functions,
such as cleaning up oil spills or delivering tumor-suppressing drugs to specific locations within the body \citep{Hasty01, Hasty02}.

\begin{figure}
\centering%
\includegraphics[width=150pt]{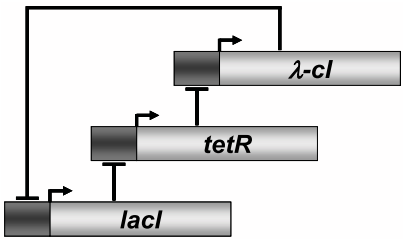}
\caption{Schematic diagram of the repressilator. Each gene (\textit{lacI\/}, \textit{tetR\/}, $\lambda$\/-\textit{cI\/}) produces a protein which binds to the
operator site of the promoter driving expression of the next gene in the sequence, thus repressing it.  Within the correct region of parameter space the
repressilator oscillates, a so-called ``ring oscillator" \citep{Hasty02}.}
\label{fig:RepSchem}
\end{figure}

Numerous artificial biological circuits have been constructed in bacteria and demonstrated to perform as designed.  One such network is the repressilator, a
three-gene synthetic genetic regulatory network developed by \citet{Elow00}.  Each gene in the repressilator produces a protein which represses the next gene
in the sequence; the protein product of the last gene represses the first gene, thus closing the loop.  This construct is known in microelectronics as a ``ring
oscillator" \citep{Hasty02}. As implemented experimentally in \textit{Escherichia coli\/} \citep{Elow00}, the repressilator consists of the genes
\textit{lacI\/}, \textit{tetR\/}, and $\lambda$-\textit{cI\/}; LacI protein represses \textit{tetR\/}, etc.\ (Fig.~\ref{fig:RepSchem}). [Standard convention is
to denote genes in italicized font beginning with a lower-case letter (e.g., \textit{tetR\/}), mRNA transcripts in non-italicized font beginning with a
lower-case letter (e.g., tetR) and proteins in non-italicized font beginning with a capital letter (e.g., TetR).]

Under the right conditions, i.e., within the correct region of parameter space, the repressilator oscillates, acting as a biological clock.  However,
determining the conditions for oscillation is nontrivial and theoretical modeling was employed to identify the appropriate design criteria \citep{Elow00}. Once
functional, a particularly interesting experimental observation was the significant fluctuations in amplitude and period exhibited by the circuit. Natural
oscillators, such as circadian clocks, do not exhibit such variability \citep{Gold02, Gonze02:PNAS, Gonze06} and subsequent modeling indicated that Nature must
employ some form of regulatory control in order to overcome the problem \citep{Barkai00, Vilar02}. The repressilator thus succeeded in providing valuable
insight regarding the design principles underlying an important biological process.

The extensive use of modeling in the design and analysis of the repressilator, as well as the highly stochastic behavior exhibited by the network, motivates
our investigation using the PLA. In Table~\ref{table:RepRxns}, we show the basic form of one-third of the repressilator model (all three genes are considered
equivalent). This corresponds to the ``stochastic" model of \citet{Elow00}. Here, all reactions are treated as elementary using simple mass-action kinetics
(i.e., rates directly proportional to the reactant population levels). Each gene is assumed to have two binding sites for repressor protein, with binding
occurring sequentially, and the unbound gene transcribes mRNA 1000 times faster than the singly- or doubly-bound gene. mRNA also translates protein
autocatalytically and mRNA and protein degrade with half-lives of 120 and 600~s, respectively \citep{Elow00}.

\begin{table}[t]
\centering%
\caption{Basic form of one-third of the full repressilator model (all three genes are considered equivalent) \citep{Elow00}. $g_x$\/ represents one of the
three repressilator genes (\textit{lacI\/}, \textit{tetR\/}, or $\lambda$-\textit{cI\/}) and $p_r$\/ the corresponding repressor protein (LacI for
\textit{tetR\/}, etc.---see Fig.~\ref{fig:RepSchem}). $m_x$\/ and $p_x$\/ represent the mRNA and protein products of $g_x$\/, respectively. All reactions are
treated using simple mass-action kinetics and all parameters with inverse molar units (M$^{-1}$\/) are divided by $N_A\Omega$ prior to runtime. $k_1,k_2$\/ are
rate constants for forward repressor binding while $k_{-1},k_{-2}$\/ are for the reverse reactions. Also shown (see text for explanation) are the ``telegraph
factor" $\gamma$\/, the ``RNA factor" $\eta$\/, and the ``protein factor" $\rho$\/ (equivalent for all genes) \citep{Sethna06, Sethna04:hints}. Here, we set
$\eta\!=\!\rho\!=\!1000$ and vary $10^{-4}\!\leq\!\gamma\!\leq\!1$.}
\begin{tabular}{@{}cll@{}} \hline\hline
    \multicolumn{2}{l}{Reaction}                                        & Parameter Value \\ \hline
    $\stackrel{\rightharpoonup}{1}$,$\stackrel{\leftharpoondown}{2}$.
         & $g_x + p_r \rightleftharpoons \{g_x{\cdot}p_r\}$                 & \hspace{-7pt}$\left\{
                                                                          \begin{array}{@{}l}
                                                                              k_1=10^9\mbox{\boldmath{$\gamma$}}/\mbox{\boldmath{$\rho$}}\ \mathrm{M}^{-1}\mathrm{s}^{-1} \\
                                                                              k_{-1}=224.0\mbox{\boldmath{$\gamma$}}\ \mathrm{s}^{-1}
                                                                          \end{array}\right.\!$ \\
    $\stackrel{\rightharpoonup}{3}$,$\stackrel{\leftharpoondown}{4}$.
         & $\{g_x{\cdot}p_r\} + p_r \rightleftharpoons \{g_x{\cdot}p_r{\cdot}p_r\}$  & \hspace{-7pt}$\left\{
                                                                          \begin{array}{@{}l}
                                                                              k_2=10^9\mbox{\boldmath{$\gamma$}}/\mbox{\boldmath{$\rho$}}\ \mathrm{M}^{-1}\mathrm{s}^{-1} \\
                                                                              k_{-2}=9.0\mbox{\boldmath{$\gamma$}}\ \mathrm{s}^{-1}
                                                                          \end{array}\right.\!$ \\
    5.   & $g_x \rightarrow g_x + m_x$                                     & $k_3=0.5$\boldmath{$\eta$} s$^{-1}$ \\
    6.   & $\{g_x{\cdot}p_r\} \rightarrow \{g_x{\cdot}p_r\} + m_x$           & $k_4=5\!\times\!10^{-4}$\boldmath{$\eta$} s$^{-1}$ \\
    7.   & $\{g_x{\cdot}p_r{\cdot}p_r\} \rightarrow \{g_x{\cdot}p_r{\cdot}p_r\}+ m_x$\,\,    & $k_5=5\!\times\!10^{-4}$\boldmath{$\eta$} s$^{-1}$ \\
    8.   & $m_x \rightarrow m_x + p_x$                                     & $k_6=0.167$\boldmath{$\rho$}/\boldmath{$\eta$} s$^{-1}$ \\
    9.   & $m_x \rightarrow \emptyset$                                   & $k_7=\ln(2)/120$ s$^{-1}$ \\
    10.  & $p_x \rightarrow \emptyset$                                   & $k_8=\ln(2)/600$ s$^{-1}$ \\
    \hline\hline
    \multicolumn{3}{l}{%
        \begin{minipage}{0pt}
            \begin{tabbing}
                \footnotesize{Initial conditions:} \\
   \hspace{6pt} \= \footnotesize{$[p_{TetR}]=0.22$~mM;}
                    \= \footnotesize{$[p_{CI}]  =2.4$~mM;}
                    \= \footnotesize{$[p_{LacI}]=0.20$~mM;} \kill
                \> \footnotesize{$[m_{TetR}]=3.8$~$\mu$\/M;}
                    \> \footnotesize{$[m_{CI}]  =8.1$~$\mu$\/M;}
                    \> \footnotesize{$[m_{LacI}]=0.15$~$\mu$\/M;} \\
                \> \footnotesize{$[p_{TetR}]=0.22$~mM;}
                    \> \footnotesize{$[p_{CI}]  =2.4$~mM;}
                    \> \footnotesize{$[p_{LacI}]=0.20$~mM;} \\
                \> \footnotesize{$g_{TetR} = g_{CI} = g_{LacI} = 1$ (molecule);} \\
                \> \footnotesize{all $\{g_x{\cdot}p_r\}$ and $\{g_x{\cdot}p_r{\cdot}p_r\}=0$.}
            \end{tabbing}
        \end{minipage}
    }
\end{tabular}
\label{table:RepRxns}
\end{table}

We also include in Table~\ref{table:RepRxns} various multiplicative factors: a ``telegraph factor" $\gamma$\/, an ``RNA factor" $\eta$\/ and a ``protein
factor" $\rho$\/ \citep{Sethna06, Sethna04:hints}.  These factors allow us to control and tune the various sources of noise in the system. For example,
increasing $\eta$\/ increases the rates of gene transcription, resulting in larger mRNA populations and less mRNA-related ``shot noise," i.e., noise arising
from the fact that the system is comprised of discrete numbers of interacting entities (in electrical circuits, shot noise arises from discrete numbers of
charge carriers; in optical devices, from discrete numbers of photons) \citep{Sethna06, Sethna04:hints}. The translation rate is divided by $\eta$\/, however,
thus cancelling out the effect of increased mRNA levels on the protein production rates. Protein-related shot noise is controlled similarly through the protein
factor $\rho$\/ while the amount of ``telegraph noise," i.e., that associated with the random switching between the ON and OFF states of the genes (reminiscent
of an electronic telegraph transmitting Morse code) \citep{Sethna06, Sethna04:hints}, is controlled through the parameter $\gamma$\/.

In this article, we focus primarily on the telegraph factor $\gamma$\/.  We do so because the performance of the leaping algorithm is strongly affected by this
parameter: at small values the method performs exceptionally well but falters as it is increased, approaching the performance of the SSA (see
Sec.~\ref{sec:Results::CaOsc}). With the system volume $\Omega\!=\!1.4\!\times\!10^{-15}$~l (the volume of a typical \textit{E.~Coli\/} cell) and $\eta$\/ and
$\rho$\/ set to high values (i.e., 1000) in order to dampen the mRNA- and protein-related noise sources, we investigate how the system behavior changes for
$10^{-4}\!\leq\!\gamma\!\leq\!1$.  We thus observe how the actual values of the gene-protein binding and unbinding rate constants, as opposed to simply their
ratios, affect the overall dynamical behavior of the system as well as the performance of the PLA.

We also find it convenient to investigate a reduced form of the repressilator model obtained by applying the ``partial equilibrium assumption" (PEA) to the
first four reactions in Table~\ref{table:RepRxns}.  Assuming each reversible reaction to be in rapid equilibrium, simple algebra leads to effective rate
expressions of the \emph{Adair\/} form \citep{Bowden04} for mRNA production from the free, singly-bound and doubly-bound genes (see Appendix~\ref{appx:Adair}
for derivations). These expressions are strictly valid in the limit $\gamma\!\rightarrow\!\infty$.  Doing so reduces the 30 reactions of
Table~\ref{table:RepRxns} to 18 in Table~\ref{table:RepReduced}. Note that the reduced model in Table~\ref{table:RepReduced} differs from the ``deterministic"
model of \citet{Elow00} in that the expressions in Table~\ref{table:RepReduced} are directly derivable from the reactions in Table~\ref{table:RepRxns} via
application of the PEA while those in Ref.~\citep{Elow00} are not.

\begin{table}[t]
\centering%
\caption{Basic form of one-third of the \emph{reduced\/} repressilator model.  Parameter values are the same as in Table~\ref{table:RepRxns}. The
\emph{Adair\/} functional forms \citep{Bowden04} describing mRNA production are similar to the well-known Hill forms, but are formally correct for
$\gamma\!\rightarrow\!\infty$ (see Appendix~\ref{appx:Adair}).}
\begin{tabular}{ll@{\hspace{20pt}}l} \hline\hline
    \multicolumn{2}{l}{Reaction}                                                            & Rate Expression \\ \hline
    1.\,\, & $g_x \rightarrow g_x + m_x$                                                    & $k_3K_1K_2/f([p_r])$ \\
    2.     & $\{g_x{\cdot}p_r\} \rightarrow \{g_x{\cdot}p_r\} + m_x$                        & $k_4K_2[p_r]/f([p_r])$ \\
    3.     & $\{g_x{\cdot}p_r{\cdot}p_r\} \rightarrow \{g_x{\cdot}p_r{\cdot}p_r\}+ m_x$\,\, & $k_5[p_r]^2/f([p_r])$ \\
    4.     & $m_x \rightarrow m_x + p_x$                                                    & $k_6[m_x]$ \\
    5.     & $m_x \rightarrow \emptyset$                                                    & $k_7[m_x]$ \\
    6.     & $p_x \rightarrow \emptyset$                                                    & $k_8[p_x]$ \\
    \hline\hline
    \multicolumn{3}{l}{\footnotesize{$K_i\!\equiv\!k_{-i}/k_i$}, ($i\!=\!1,2$)} \\
    \multicolumn{3}{l}{\footnotesize{$f([p_r])\!\equiv\!K_1K_2+K_2[p_r]+[p_r]^2$}} \\
\end{tabular}
\label{table:RepReduced}
\end{table}

\section{Methods} \label{sec:Methods}

To carry out our investigations, we use the partitioned leaping algorithm \citep{Harris06}, an extension and variant of the $\tau$\/-leaping method of
Gillespie and co-workers \citep{Gillesp00, Gillesp01, Gillesp03, Cao06:newStep}. The PLA operates on the same basic principles that underlie all leaping
methods: calculate a time step $\tau$\/ over which all reaction rates in the system remain ``essentially constant" and then determine the number of times each
reaction fires within that interval by sampling from an appropriate probability distribution. The primary difference between our approach and other leaping
algorithms is that we utilize the \emph{entire\/} theoretical framework developed by Gillespie \citep{Gillesp00, Gillesp01} for bridging from the
discrete-stochastic description of reaction dynamics to the more familiar continuous-deterministic representation \citep{Harris06}.

At each step of a PLA simulation, reactions are partitioned, using theoretically-sound criteria \citep{Gillesp00, Gillesp01}, into four categories based on the
calculated time step and the current reactant population levels. The categories correspond to different levels of approximation; reactions with small reactant
populations garner a detailed ``exact-stochastic" classification (i.e., a SSA treatment) while those with larger populations receive coarser descriptions. The
coarse classifications range from ``Poisson" to ``Langevin" to ``deterministic," with the levels of stochasticity decreasing and approximation increasing as
one moves up the hierarchy. The result is a truly multiscale method, where fluctuations associated with rare events are correctly described while frequent
events ``leapt" over multiple reaction firings at a time. As such, the PLA accomplishes what so-called hybrid methods \citep{Hasel02, Kiehl04, Taka04, Vasu04,
Burr04, Puch04, Salis05:hybrid, Griff06, Wylie06} aim to do but in a much more simple and theoretically-sound way. For even moderately-sized systems the
computational gains of the PLA relative to the SSA can be significant \citep{Harris06}.  We refer the interested reader to Ref.~\citep{Harris06} for further
details regarding the theoretical foundations and implementation of the PLA.

\begin{figure}
\includegraphics*[width=240pt]{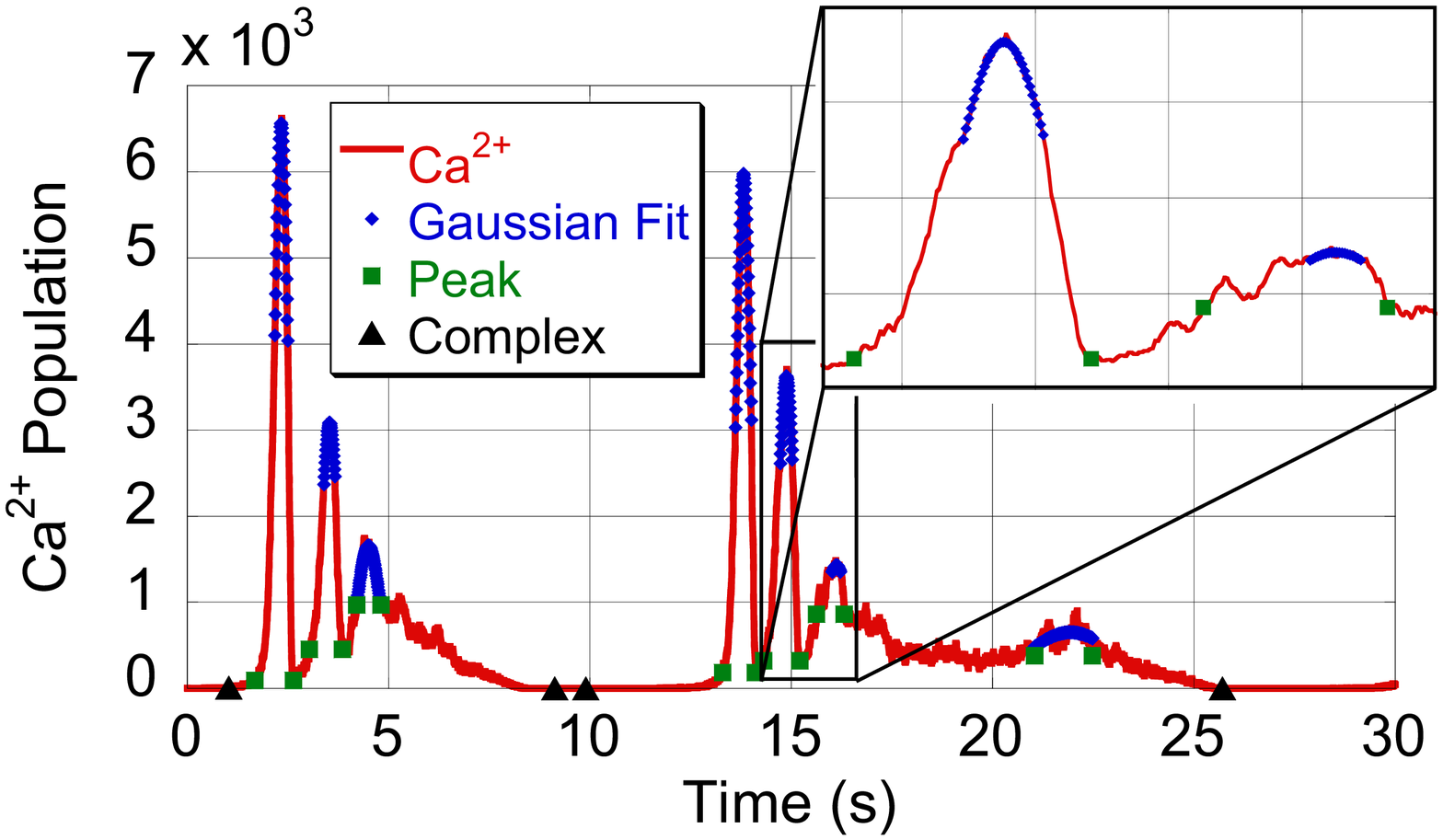}
\caption{Example calcium-oscillations time course and the Gaussian fits obtained using the peak-analysis software employed in this work. Results are for a
system volume $\Omega\!=\!10^{-21}$~l. Also shown are peak and peak-complex bracket points identified by the fitting algorithm. Notice that stochastic effects
lead at this small volume to the identification of a fourth peak in the second peak complex. (\textit{Inset}) Blown up view of the second and third peaks in
the second peak complex.  Squares correspond to where fitting began, diamonds to where fitting concluded.}
\label{fig:peakFits}%
\end{figure}

The promise of the PLA, and leaping algorithms in general, is that long-time stochastic simulations can be performed, allowing for large-scale data collection
and quantitative statistical analyses of the resulting time series. However, leaping simulations produce noisy time-evolution trajectories and automated data
collection tools that can compensate for noise are needed. Here, we use in-house time-domain peak-analysis software for this purpose. Borrowing ideas from the
automated identification of peaks in mass spectral data \citep{Wall04, Kears05}, the software identifies ``significant" peaks within a time series and fits
Gaussians to the data in order to wash out the noise.  An example calcium-oscillations time series and the Gaussian fits achieved using the peak-analysis
software are shown in Fig.~\ref{fig:peakFits}.

Using this tool, we collect large amounts of peak amplitude and peak-to-peak distance data from simulated time series and perform various statistical analyses.
We calculate averages and variances from long-time PLA and deterministic simulation runs (see \cite{note:determVar} for an explanation of variance in the
deterministic case) and perform z-tests on the differences in means and F-tests on the ratios of variances \citep{IntroStats}. We also calculate coefficients
of variation (COVs), defined as the ratio of the standard deviation to the mean \citep{Kaern05}, in order to quantify the relative \emph{importance\/} of the
noise. Finally, we put the data into the form of smoothed histograms \citep{Harris06} and calculate ``histogram distances," $D$\/, and ``self distances,"
$D^\mathrm{self}_\mathrm{Ref}$ \citep{Cao06:histDist, Harris06}, so as to account for any particulars in the \emph{shapes\/} of the distributions (e.g., long
tails, bimodal features, etc.). We do all of this for various system properties (i.e., volumes, telegraph factors) in order to quantify changes in the system
behavior and to identify points of transition to determinism.

All PLA simulations reported in this work were performed using the parameters `$\approx\!{1}$'=3, `$\gg\!{1}$'=100 \citep{Harris06} and using the species-based
$\tau$\/-selection procedure of \citet{Cao06:newStep}, as modified in \citet{Harris06}, with ``error control parameter" $\epsilon\!=\!0.03$. These represent
``typical" values for these parameters. Derivations of the $g_i$\/ values \citep{Cao06:newStep, Harris06} used in $\tau$\/~selection for the Michaelis-Menten
and Adair reactions of Tables~\ref{table:CaRxns} and \ref{table:RepReduced} are given in Appendix~\ref{appx:tauSelec}. It should also be noted that an
attractive feature of the PLA is its ability, via simple manipulation of the classification parameters, to force simulations at any level of description
\citep{Harris06}. Thus, both deterministic and exact-stochastic simulations reported in this work were performed using the same code as for the PLA
simulations. The PLA segues to an explicit Euler method in the deterministic limit and to the next-reaction method \citep{Gibson00} in the exact-stochastic
limit.

\section{Results} \label{sec:Results}

\subsection{Calcium Oscillations} \label{sec:Results::CaOsc}

\begin{figure*}
\includegraphics*[width=350pt]{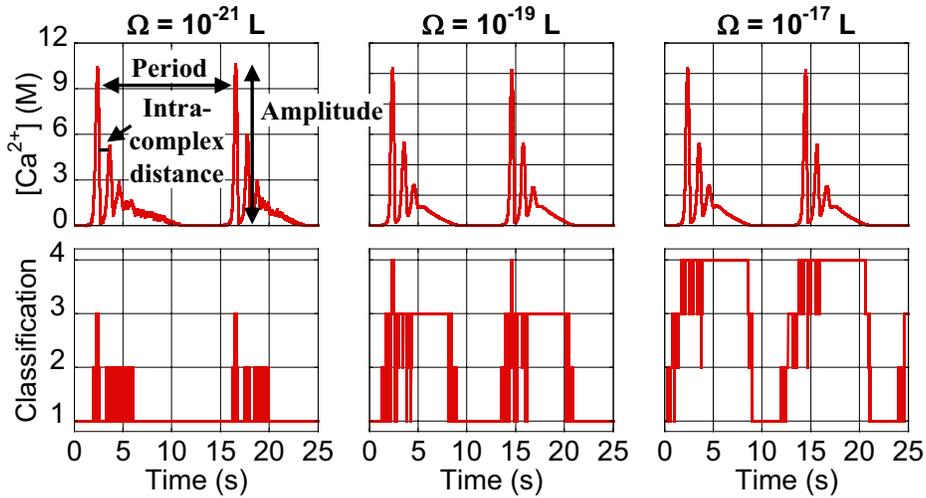}
\caption{Example Ca$^{2+}$ periodic-bursting time courses and associated classifications for $G_\alpha\!+\!\mathrm{Ca}\!\rightarrow\!\mathrm{Ca}$\/
(Table~\ref{table:CaRxns}, reaction~4) obtained using the PLA at three different system volumes. Classifications are: (1) exact stochastic, (2) Poisson, (3)
Langevin, (4) deterministic. Also shown (\textit{top-left panel\/}) are the three system attributes investigated: first-peak amplitudes, first-to-second
intra-complex distances and first-to-first inter-complex periods.}
\label{fig:CaPopsClassif}%
\end{figure*}

The periodic-bursting regime of the \citet{Kumm00} calcium-oscillations model (Table~\ref{table:CaRxns}) is characterized by large-amplitude complex
oscillations in which the Ca$^{2+}$ repeating unit is a three-peak complex.  In Fig.~\ref{fig:CaPopsClassif}, we show example time courses at three different
system volumes spanning four orders of magnitude obtained using the PLA. Also shown are the classifications achieved along the time courses for the reaction
$G_\alpha+\mathrm{Ca}\rightarrow\mathrm{Ca}$\/ (Table~\ref{table:CaRxns}, reaction~4). The classifications range from 1--4, with 1 being the finest level of
description (exact stochastic) and 4 the coarsest (deterministic).

The plots in Fig.~\ref{fig:CaPopsClassif} starkly illustrate why this system is ideally suited for treatment via the PLA: \emph{the classifications oscillate
in time along with the reactant species populations\/}. When the Ca$^{2+}$ population is small we see that the reaction gets classified at the exact-stochastic
level, while coarser descriptions are employed when the population is large (similar behavior is seen for other reactions in the system as well---data not
shown). As such, the PLA is able to accurately capture stochastic effects that arise in this system when the species populations become small without suffering
from the characteristic inefficiency of the SSA when the populations become large.

\begin{figure}
\centering%
\includegraphics[width=175pt]{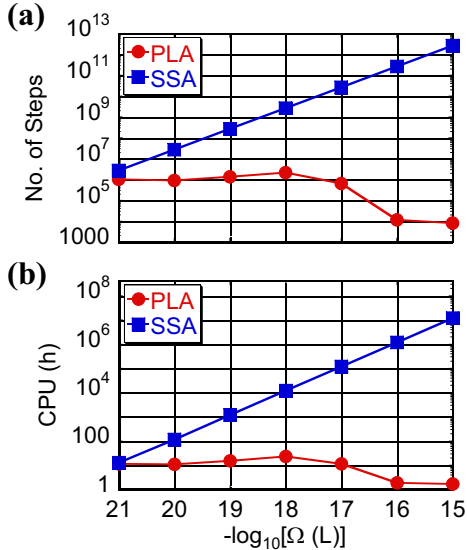}
\caption{Average numbers of steps from (\emph{a\/}) and total CPU times for (\emph{b\/}) $10\,000$ PLA and SSA simulation runs of 20~s for the \citet{Kumm00}
core model for calcium oscillations (Table~\ref{table:CaRxns}).  Both graphs are shown as log-log plots.  SSA values at $\Omega\!=\!10^{-20}$ and $10^{-19}$~l
are based on 1000 and 100 simulation runs, respectively.  SSA values at $\Omega\!\ge\!10^{-18}$~l are extrapolations (not based on actual data).  Note that the
PLA steps and CPU times go through maxima at $\sim\!\Omega\!=\!10^{-18}$~l. Similar behavior was observed for the example systems in Ref.~\citep{Harris06}.
Also note that in the case of the SSA, the linear relationship between computational expense and system size \citep{Gillesp76, Gillesp77, Gillesp07}, which has
the form $y\!=\!mx$\/, with $m$\/ being the slope (the y-intercept is zero since, obviously, a system of zero size requires zero computational effort), appears
here as a line with a slope of unity and y-intercept of $\log_{10}(m)$. All simulations were performed on a 3.60~GHz Pentium Xeon processor.}
\label{fig:CaTimings}
\end{figure}

This is evident in Fig.~\ref{fig:CaTimings}, where we show results of a step and timing analysis comparing the performance of the PLA to the SSA. As expected,
we see a linear increase in the computational expense of the SSA with increasing system size (see Fig.~\ref{fig:CaTimings}, caption) \citep{Gillesp76,
Gillesp77, Gillesp07}. The PLA, on the other hand, exhibits more complex behavior, with the expense initially remaining constant, then increasing slightly,
going through a maximum at $\sim\!\Omega\!=\!10^{-18}$~l and then dropping off sharply before finally leveling off. Interestingly, similar behavior was seen
for the simple example systems in Ref.~\citep{Harris06}, which were specifically designed to showcase the strengths of the algorithm. Most importantly,
however, is that Fig.~\ref{fig:CaTimings} clearly illustrates that for all but the smallest system size considered the PLA \emph{far\/} outperforms the SSA, by
as many as eight orders of magnitude in simulation steps at $\Omega\!=\!10^{-15}$~l. It is these types of accelerations that make quantifying stochastic
effects in this system possible, something that was unachievable in Ref.~\citep{Kumm05} because of the limitations of the SSA.

\begin{figure*}
\centering%
\includegraphics*[width=500pt]{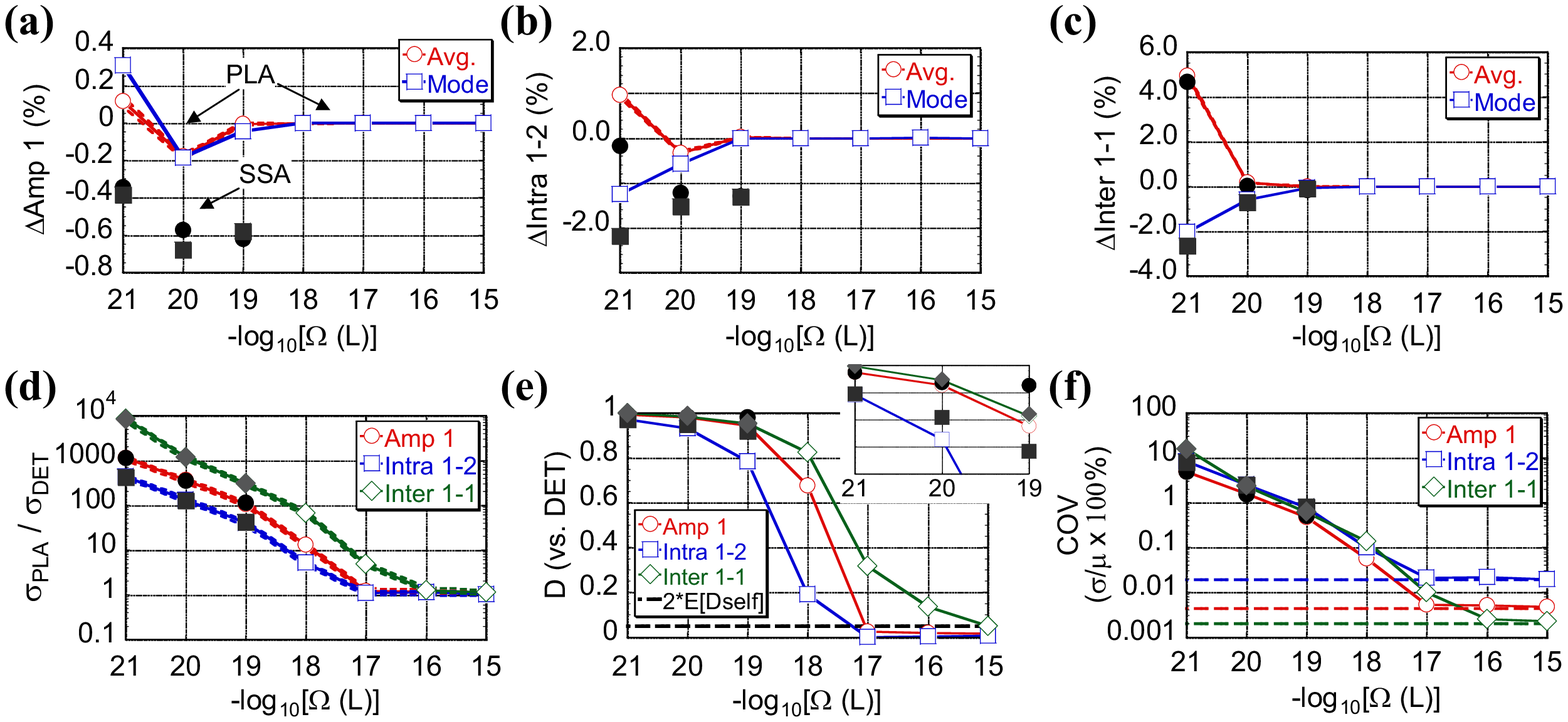}
\caption{Statistical results for the Ca$^{2+}$ periodic-bursting regime. Results of PLA and SSA simulations are compared to deterministic (DET) predictions.
PLA results are shown as empty symbols (circle, square, diamond) connected by lines.  SSA results are shown as disconnected, filled symbols. PLA and SSA points
designated with the same symbol correspond to the same quantity [e.g., in (a), the PLA averages are given as empty circles while the SSA averages are shown as
filled circles]. All PLA and deterministic values are based on over $10\,000$ collected data points. For the SSA, over $10\,000$ data points were collected for
$\Omega\!=\!10^{-21}$ and $10^{-20}$~l and $\sim\!1000$ were collected for $\Omega\!=\!10^{-19}$~l.  No SSA results are given for $\Omega\!\ge\!10^{-18}$~l due
to computational expense.
(\emph{a\/})--(\emph{c\/}): Deviations from determinism, shown as percentages [(\{PLA or SSA\}$-$DET)$\div$DET$\times100\%$], for averages and modes of
Ca$^{2+}$ first-peak amplitudes (Amp~1), first-to-second intra-complex distances (Intra~1-2), and first-to-first inter-complex periods (Inter~1-1),
respectively (see Fig.~\ref{fig:CaPopsClassif}, \emph{top-left panel\/}). Dashed lines denote 95\% confidence intervals on the PLA averages [difficult to see
in (b) and (c)]. Note that long-tailed distributions lead to averages and modes on opposite sides of the deterministic predictions at small volumes in (b) and
(c).
(\emph{d\/}): Ratios of standard deviations (\{PLA or SSA\}$\div$DET) for the three attributes in (a)--(c). Dashed lines denote 80\% confidence intervals
(because of the relative weakness of the F-test \citep{IntroStats}).
(\emph{e\/}): Histogram distances (\{PLA or SSA\} vs.\ DET) for the three attributes in (a)--(c). The dashed line denotes \emph{twice\/} the deterministic self
distance ($2\!\times\!\mathrm{E}[D^\mathrm{self}_\mathrm{DET}]$) (see \cite{note:selfDist}). The self distances for all three attributes are essentially
identical in this case.
(\emph{f\/}): Coefficients of variation (COVs) obtained from PLA and SSA simulations, shown as percentages (standard deviation$\div$average$\times100\%$), for
the three attributes in (a)--(c).  Deterministic limits are given as dashed lines.}
\label{fig:CaStats}
\end{figure*}

Our statistical results are shown in Fig.~\ref{fig:CaStats}.  In all cases, we compare results obtained from both PLA and SSA simulations to deterministic
predictions for the three attributes considered, namely, first-peak amplitudes, first-to-second intra-complex distances, and first-to-first inter-complex
periods (see Fig.~\ref{fig:CaPopsClassif}, \emph{top-left panel\/}). In the case of the SSA, we were only able to obtain data for the three smallest system
sizes considered because of the computational expense of the method.

In Figs.~\ref{fig:CaStats}\textit{a\/}--\ref{fig:CaStats}\textit{c\/}, we compare averages and modes obtained from the PLA and SSA to deterministic
predictions.  The results are shown as percent deviations from determinism.  In all cases, we see small yet statistically significant deviations from
determinism at small volumes and, in the case of the PLA, a rapid convergence to the deterministic limit with increasing system size. Close inspection reveals
that full convergence is achieved for all attributes by $\Omega\!=\!10^{-18}$~l.  It is also clear in Figs.~\ref{fig:CaStats}\textit{a\/} and
\ref{fig:CaStats}\textit{b\/} that there are discrepancies between the PLA results and the SSA results.  The discrepancies are small, however, on the order of
1\% or less in all cases, and decrease with decreasing $\epsilon$\/ (data not shown). Interestingly, there are virtually no discrepancies between the PLA
results and the SSA results in Fig.~\ref{fig:CaStats}\textit{c\/}, the inter-complex periods.  We cannot at present explain why the PLA achieves greater
accuracy for this attribute over the others. Understanding the sources of error in leaping algorithms and developing strategies for attenuating them is an area
of current interest \cite{Rath03, Cao04:Stability}. Suffice it to say that in this case the PLA achieves very good to excellent accuracy for all quantities
considered.

In Figs.~\ref{fig:CaStats}\textit{d\/} and \ref{fig:CaStats}\textit{e\/}, we consider the \emph{distributions\/} of the attributes.
Figure~\ref{fig:CaStats}\textit{d\/} shows data for standard deviations, a simple point statistic, while Fig.~\ref{fig:CaStats}\textit{e\/} considers the
shapes of the distributions through the histogram distance \citep{Cao06:histDist, Harris06}.  In Fig.~\ref{fig:CaStats}\textit{d\/}, we see almost perfect
correspondence between the PLA and the SSA results.  In Fig.~\ref{fig:CaStats}\textit{e\/}, however, we see discrepancies in the histogram distances for the
amplitude and the intra-complex distance (see inset).  Taken together, along with Figs.~\ref{fig:CaStats}\textit{a\/} and \ref{fig:CaStats}\textit{b\/}, this
indicates that the PLA is accurately capturing the shapes of the distributions but they are shifted slightly relative to those obtained with the SSA.

As far as convergence to determinism, both Figs.~\ref{fig:CaStats}\textit{d\/} and \ref{fig:CaStats}\textit{e\/} give the same result: \emph{the different
attributes converge to the deterministic limit at different rates and with different transition points\/}. The intra-complex distance converges the fastest,
followed by the peak amplitude and finally the inter-complex period.  The amplitude and intra-complex distance statistically converge to the deterministic
limit at $\Omega\!=\!10^{-17}$~l while the period converges at $10^{-15}$~l.  These convergence points differ from those for the averages by one to three
orders of magnitude (cf.\ Figs.~\ref{fig:CaStats}\textit{a\/}--\ref{fig:CaStats}\textit{c\/}) and indicate a persistence of noise in this system at volumes
much larger than expected based on the analysis of Ref.~\citep{Kumm05}.

Finally, in Fig.~\ref{fig:CaStats}\textit{f\/} we consider the relative ``importance" of the noise through the coefficient of variation (COV).  The idea is
that even if noise in an attribute is significant from a statistical perspective it might be so subtle as to be of little practical import. For example, in
this case we see that for $\Omega\!\ge\!10^{-20}$~l the COVs for all attributes are less than a few percent (the discrepancies between the PLA and the SSA seen
in Figs.~\ref{fig:CaStats}\textit{a\/} and \ref{fig:CaStats}\textit{b\/} are virtually indiscernible on this scale).  The noise effects clearly persist up
until $10^{-17}$~l (as seen in Figs.~\ref{fig:CaStats}\textit{d\/} and \ref{fig:CaStats}\textit{e\/} as well) but it seems unlikely that in any realistic
setting, e.g, an embedding within a larger ``whole-cell" model, they would be of much practical consequence. Whether or not this is true (it is debatable
\citep{Samo06:Dev, Samo06:STKE}), it is certainly the case that it would be difficult, if not impossible, to perceive these effects visually. This explains,
therefore, why \citet{Kumm05} reported the stochastic-to-deterministic transition point for this model to be at $\sim\!10^{-20}$~l (tens of thousands of
Ca$^{2+}$ ions).  Our results thus largely corroborate their claim that a deterministic treatment is justified for volumes larger than this.

\subsection{Repressilator} \label{sec:Results:Repress}

Our analysis of the repressilator focuses on behavioral changes that arise when the intermittent rates of switching between the transcriptional ON and OFF
states of the genes are varied.  The parameter that controls this is the telegraph factor $\gamma$\/. From an intuitive standpoint, we expect to observe large
deviations from determinism at small values of $\gamma$\/ and a convergence towards deterministic behavior with increasing $\gamma$\/ because of the
``averaging out" of the states of the genes \citep{Kaern05}. Moreover, by making the RNA and protein factors, $\eta$\/ and $\rho$\/, large we minimize the
effects of shot noise. However, we cannot eliminate it completely, and thus we expect to encounter some residual effects. Finally, we also expect that the PLA
simulations will begin to bog down as $\gamma$\/ is increased because of the growing disparities between the gene-protein binding and unbinding rates and the
rates of all other reactions in the system \citep{Harris06}.

\begin{figure}
\centering%
\includegraphics[width=240pt]{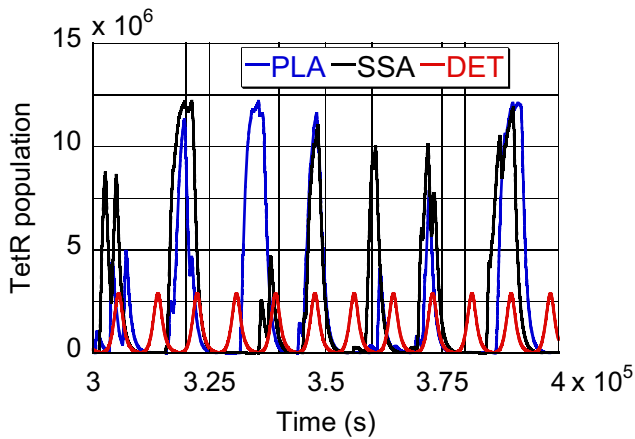}
\caption{Example time courses (TetR protein) illustrating ``deviant effects" \citep{Samo06:Dev} in the repressilator at small values of $\gamma$\/. With
$\gamma\!=\!10^{-4}$ and $\eta\!=\!\rho\!=\!1000$, stochastic realizations (PLA and SSA) differ markedly from the deterministic prediction.}
\label{fig:RepDeviant}
\end{figure}

\begin{figure*}
\centering%
\includegraphics[width=500pt]{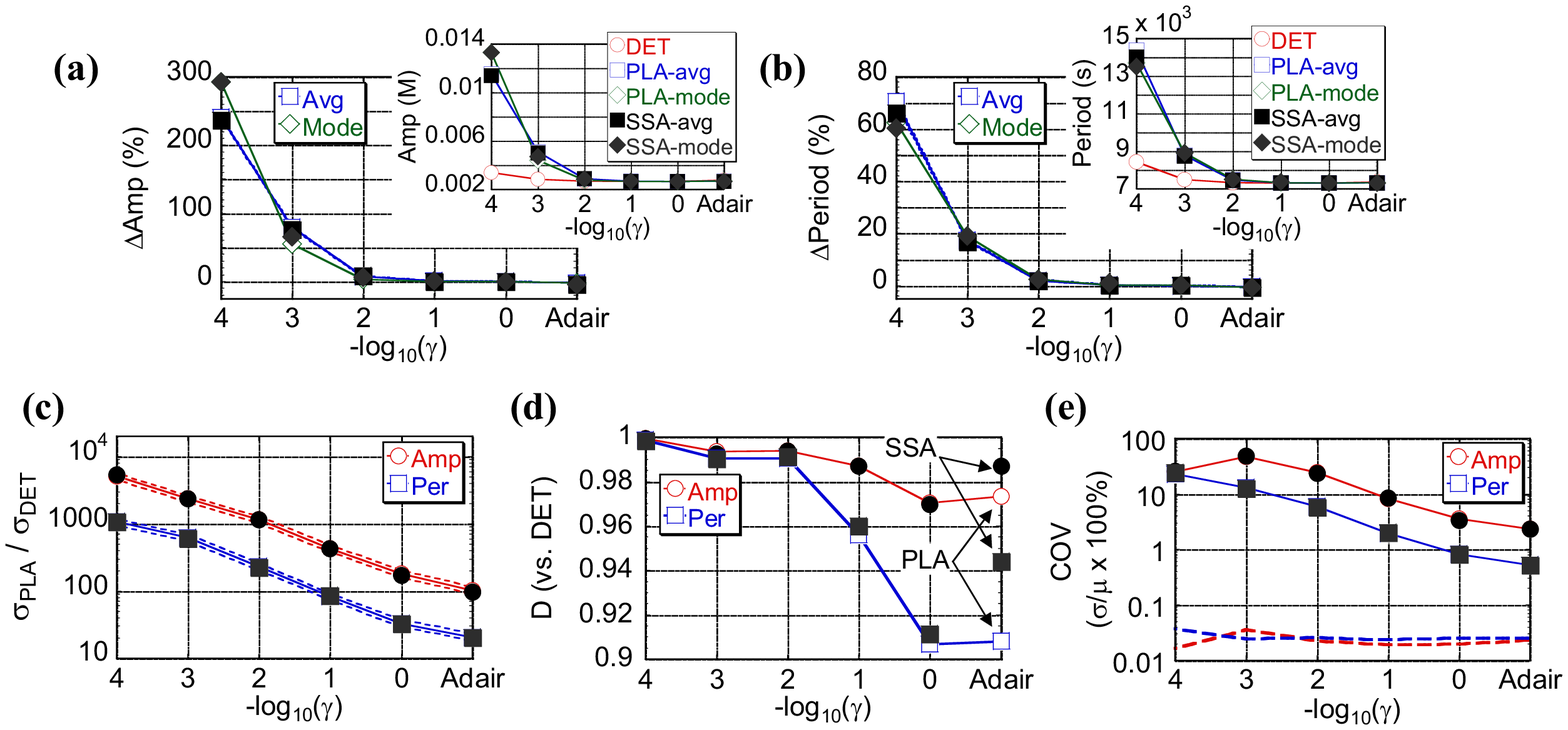}
\caption{Statistical results for the repressilator.  At various values of the telegraph factor $\gamma$\/, and at the Adair limit
($\gamma\!\rightarrow\!\infty$), results of both PLA and SSA simulations are compared to deterministic (DET) predictions. As in Fig.~\ref{fig:CaStats}, PLA
(and deterministic) results are shown as empty symbols connected by lines, SSA results are shown as disconnected, filled symbols, and PLA and SSA points
designated with the same symbol correspond to the same quantity.  All PLA and deterministic values are based on over $10\,000$ collected data points while all
SSA values are based on $\sim\!1000$ collected data points. Note that the only clearly discernible discrepancies between the PLA and SSA results are the
histogram distances in (d) at the Adair limit.
(\emph{a\/}),(\emph{b\/}): Averages and modes for the TetR-protein peak amplitude and period, respectively.  In the main plots, results are given as percent
deviations from determinism (95\% confidence intervals on the PLA averages are difficult to see). In the insets, results are shown in absolute form,
illustrating the dependencies of the amplitude and period on $\gamma$\/.
(\emph{c\/}): Ratios of standard deviations (\{PLA or SSA\}$\div$DET) for the TetR-protein peak amplitude and period. Dashed lines denote 80\% confidence
intervals.
(\emph{d\/}): Histogram distances (\{PLA or SSA\} vs.\ DET). Note that the self distances are off the chart.
(\emph{e\/}): Coefficients of variation, given as percentages, obtained from both PLA and SSA simulations. In principle, the deterministic limits (dashed
lines) vary with $\gamma$\/ [see (a) and (b), \textit{insets\/}], though here they are very nearly constant.}
\label{fig:RepStatsDet}
\end{figure*}

\begin{figure}
\centering%
\includegraphics[width=175pt]{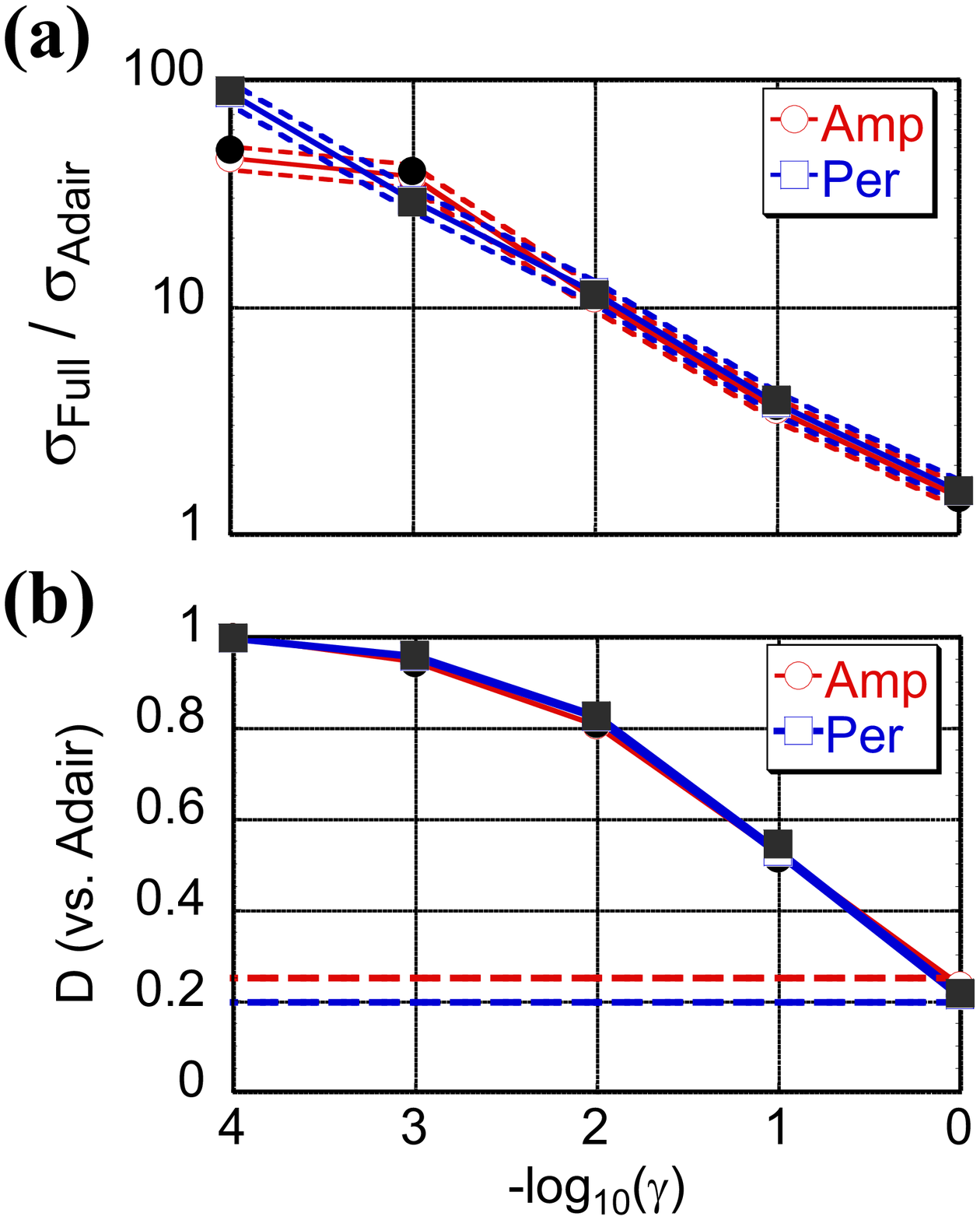}
\caption{Convergence of the full repressilator model to the Adair limit with increasing $\gamma$\/.  At each value of $\gamma$\/, PLA and SSA results of the
full model (Table~\ref{table:RepRxns}) are compared to PLA and SSA results, respectively, of the reduced model (Table~\ref{table:RepReduced}).  In all cases,
the PLA and SSA values (empty and filled symbols, respectively) coincide almost perfectly. (\textit{a\/}): Ratios of standard deviations (full$\div$reduced)
for the TetR-protein peak amplitude and period. Dashed lines denote 80\% confidence intervals. (\textit{b\/}): Histogram distances (full vs.\ reduced). Dashed
lines denote twice the \emph{Adair\/} self distances.}
\label{fig:RepStatsAdair}
\end{figure}

\begin{figure}
\centering%
\includegraphics[width=175pt]{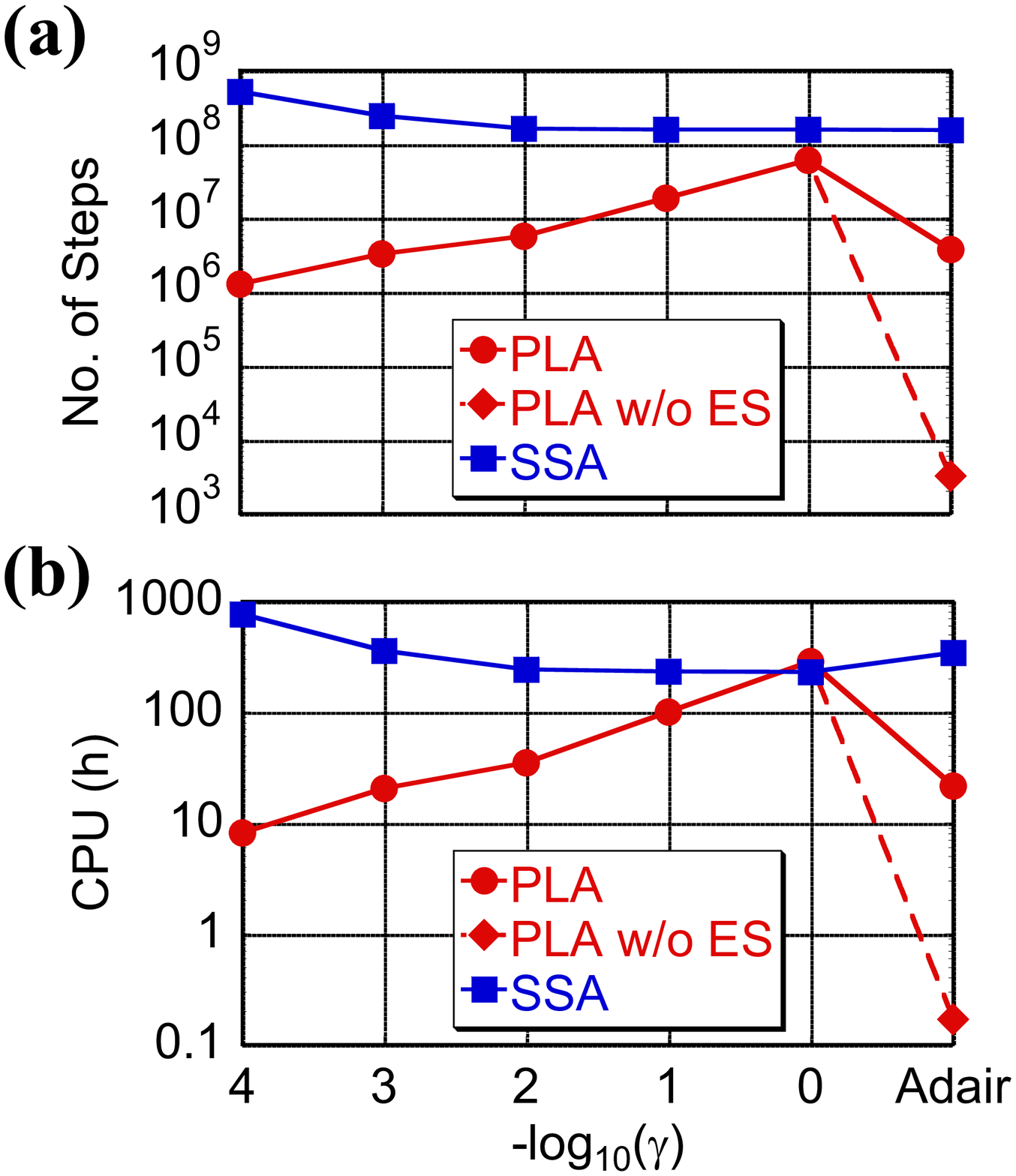}
\caption{Average numbers of steps from (\emph{a\/}) and total CPU times for (\emph{b\/}) 1000 PLA and SSA simulation runs of $30\,000$~s of the full
(Table~\ref{table:RepRxns}) and reduced (Table~\ref{table:RepReduced}) repressilator models.  All SSA points are based on 100 simulation runs (due to
computational expense). Note that the CPU curves in (b) cross at $\gamma\!=\!1$ because each PLA step is more expensive than each SSA step.  At the Adair
limit, results are given for PLA simulations that both include and exclude the exact-stochastic (ES) classification (see \cite{note:AdairTimings}). All
simulations were performed on a 3.60~GHz Pentium Xeon processor.}
\label{fig:RepTimings}
\end{figure}

In Figs.~\ref{fig:RepDeviant}--\ref{fig:RepTimings}, these expectations are realized.  In Fig.~\ref{fig:RepDeviant}, we show example time courses for TetR
protein (taken as representative of the system behavior) that illustrate how ``deviant effects" \citep{Samo06:Dev} arise in the repressilator at small values
of $\gamma$\/.  With $\gamma\!=\!10^{-4}$, we see in Fig.~\ref{fig:RepDeviant} that the true behavior of the system, as captured by both the PLA and the SSA,
differs markedly from that predicted deterministically. Rather than emitting smooth and regular oscillations, the system produces large-amplitude intermittent
``bursts" of (mRNA and) protein production.  This is a direct consequence of the slow stochastic switching between the ON and OFF states of the genes and is
consistent with gene-expression behavior often observed in eukaryotes \citep{Kaern05, Blake06}. Note that due to stochasticity the PLA and SSA traces differ
from each other.  As we shall see, however, they are virtually identical from a statistical standpoint.

In Fig.~\ref{fig:RepStatsDet}, we present results of our statistical analyses of the repressilator.  At various values of $\gamma$\/, as well as at the Adair
limit ($\gamma\!\rightarrow\!\infty$), we compare the stochastic behavior of the system, as captured by both the PLA and the SSA, to deterministic predictions.
In Figs.~\ref{fig:RepStatsDet}\textit{a\/} and \ref{fig:RepStatsDet}\textit{b\/}, we consider averages and modes for the TetR-protein peak amplitude and
period, respectively.  In both cases, the PLA and SSA results coincide almost perfectly and show large deviations from determinism at small values of
$\gamma$\/ and a convergence towards the deterministic limit with increasing $\gamma$\/.  Close inspection of the PLA results reveals that statistical
convergence to the deterministic limit is achieved for both attributes by $\gamma\!=\!1$. It is also evident from these plots that the behavior of the full
model (Table~\ref{table:RepRxns}) approaches that of the reduced model (Table~\ref{table:RepReduced}) with increasing $\gamma$\/, as we would expect.

In Figs.~\ref{fig:RepStatsDet}\textit{c\/} and \ref{fig:RepStatsDet}\textit{d\/}, we consider the distributions of the amplitude and the period.  Again, we
look at ratios of standard deviations and histogram distances and again we see a convergence towards determinism with increasing $\gamma$\/. However, in this
case the deterministic limit is never reached; even at the Adair limit we see considerable deviation from determinism.  Furthermore, we see very good
correspondence between the PLA and the SSA results.  In fact, the only significant differences that we see are the small discrepancies in the histogram
distances at the Adair limit in Fig.~\ref{fig:RepStatsDet}\textit{d\/}. This is interesting in light of the discrepancies seen between the PLA and the SSA in
Figs.~\ref{fig:CaStats}\textit{a\/} and \ref{fig:CaStats}\textit{b\/} for the calcium-oscillations model, which also contains reduced reaction types (see
Table~\ref{table:CaRxns}).  This suggests that reduced reactions might be the source of the various inaccuracies seen in Figs.~\ref{fig:CaStats} and
\ref{fig:RepStatsDet}. We plan to investigate this issue further in the future.

In Fig.~\ref{fig:RepStatsDet}\textit{e\/}, we consider the noise strength through the COV.  Here, as in Figs.~\ref{fig:RepStatsDet}\textit{c\/} and
\ref{fig:RepStatsDet}\textit{d\/}, we see almost perfect agreement between the PLA and the SSA results and an incomplete convergence towards the deterministic
limit with increasing $\gamma$\/. It is clear, therefore, that significant shot noise effects persist in this system even as $\gamma\!\rightarrow\!\infty$\/.
Moreover, it is interesting to note the elevated levels of noise in the amplitude as compared to the period.  We see an approximately order-of-magnitude
difference in the COVs for these two attributes at all values of $\gamma\!>\!10^{-4}$ and at the Adair limit. Contrast this with
Fig.~\ref{fig:CaStats}\textit{f\/}, which shows no appreciable difference between the COVs for the amplitude and the period in the calcium-oscillations model.
This is an example of the type of fine-level insight that we can garner via the leaping algorithm.

It is clear from Figs.~\ref{fig:RepStatsDet}\textit{c\/}--\ref{fig:RepStatsDet}\textit{e\/} that the repressilator never behaves in a fully deterministic
manner under the conditions that we consider. However, it is also clear that the behavior does approach that of the reduced model with increasing $\gamma$\/.
Therefore, in Fig.~\ref{fig:RepStatsAdair} we quantify this convergence to the Adair limit by repeating the statistical tests of
Figs.~\ref{fig:RepStatsDet}\textit{c\/} and \ref{fig:RepStatsDet}\textit{d\/} but using the PLA and SSA results for the reduced model, rather than the
deterministic results at each $\gamma$\/, as our reference. The results clearly confirm the (near) convergence of the system behavior to the Adair limit at
$\gamma\!=\!1$.

Finally, in Fig.~\ref{fig:RepTimings} we present results of a step and timing analysis comparing the performance of the PLA to the SSA for simulations of both
the full (Table~\ref{table:RepRxns}) and reduced (Table~\ref{table:RepReduced}) repressilator models. For the full model, we see the convergence in
computational expense of the PLA and the SSA that we anticipated \citep{Harris06}. In Fig.~\ref{fig:RepTimings}\textit{a\/}, the numbers of steps required for
PLA and SSA simulations converge asymptotically with increasing $\gamma$\/.  In Fig.~\ref{fig:RepTimings}\textit{b\/}, we see a similar trend for the CPU
times, although interestingly the curves here cross at $\gamma\!=\!1$ because each PLA step is more computationally expensive than each SSA step. Also worth
noticing is that both plots indicate that the expense of the SSA \emph{decreases\/} with increasing $\gamma$\/ while the opposite is true for the PLA. This is
because the protein (and mRNA) populations, which are the prime bottleneck for the SSA, tend to be larger at small values of $\gamma$\/ (cf.\
Figs.~\ref{fig:RepDeviant} and \ref{fig:RepStatsDet}\textit{a\/}).  Leaping algorithms are not affected by population sizes, however, having been developed
specifically to cope with this problem \citep{Gillesp00, Gillesp01}. Hence, we see that when stochastic effects in this system are most pronounced (small
$\gamma$\/) the PLA far outperforms the SSA.

In Ref.~\citep{Harris06}, it was posited that large disparities in rate constants would prove to be the prime bottleneck for leaping algorithms.  This is
confirmed in Fig.~\ref{fig:RepTimings} by the declining performance of the PLA with increasing $\gamma$\/. It is for exactly this reason that we consider the
reduced model of Table~\ref{table:RepReduced}.  In Figs.~\ref{fig:RepStatsDet} and \ref{fig:RepStatsAdair}, we have seen that the behavior of the full model
approaches that of the reduced model with increasing $\gamma$\/.  Now, in Fig.~\ref{fig:RepTimings} we see that the performance of the PLA is greatly enhanced
by the model reduction. Depending on exactly how we choose to implement the PLA (see \cite{note:AdairTimings} for an explanation), we can achieve gains of
between one and four orders of magnitude in both simulation steps and run times. Additionally, it is important to note that reducing the model has very little
effect on the performance of the SSA. In fact, we see in Fig.~\ref{fig:RepTimings} that while the numbers of simulation steps required for the SSA remain
virtually unchanged upon reducing the model, the CPU time actually \emph{increases\/} by $\sim\!{50}$\% because of the higher complexity rate expressions in
Table~\ref{table:RepReduced} which impose additional computational burdens on the algorithm. Our results indicate, therefore, that there is a distinct
advantage to using model reduction in conjunction with leaping which is absent with regards to the SSA.

\section{Discussion and Conclusions} \label{sec:Discuss}

Using the scarcity of published practical applications of leaping as a backdrop, our aim in this article was to investigate the performance characteristics of
a particular leaping algorithm, the PLA, when applied to two non-trivial biological models under a variety of conditions.  Our hope was to identify the
conditions under which leaping proves particularly beneficial and where it falters and, hence, provide a kind of guide that will aid and motivate future
applications of the method. Our use of the PLA, as opposed to a different leaping algorithm, was based purely on expediency, given that we developed the method
\citep{Harris06}. However, the PLA operates on the same basic principles as all leaping algorithms and its performance can thus be seen as generally reflective
of the entire class of method.  In other words, the accelerations that we have reported here are not wholly unique to the PLA but are characteristic of the
leaping methodology in general.  Similarly, the bottlenecking that we experienced in the face of fast-reversible reactions in the repressilator system can be
expected to afflict all leaping algorithms.


That being said, there are attributes of the PLA that we believe set it apart from its various counterparts, and we would be remiss in not emphasizing these.
Foremost among these is its simplicity of implementation and ease of use.  The algorithm is concise, straightforward and overcomes various technical
difficulties (e.g., negative populations \citep{Tian04, Chatt05, Cao05:negPop}) without the need for extensive auxiliary machinery \citep{Harris06}. Using the
PLA requires little more than a system definition (reactions), rate expressions (elementary or non-elementary) and definition of three simple model-independent
parameters [`$\approx\!1$', `$\gg\!1$', `$\ll\!1$' (i.e., $\epsilon$\/)] \cite{Harris06}.
%
%
Also significant is the ability to force the algorithm to perform both deterministic and exact-stochastic simulations by simple manipulation of the
classification parameters (e.g., setting `$\approx\!1$'$=\!\infty$ or `$\gg\!1$'$=\!0$).  In our case, this significantly simplified the noise quantification
and step and timing analyses.

Tangibly speaking, our results clearly illustrate the great potential that leaping methods hold in computational systems biology. For both the
calcium-oscillations model and the repressilator, we observed orders-of-magnitude accelerations relative to the SSA (Figs.~\ref{fig:CaTimings} and
\ref{fig:RepTimings}) that made quantifying stochastic effects in these systems possible. In the calcium-oscillations case, this gave us access to subtle
effects of stochasticity that would have been indiscernible otherwise (Fig.~\ref{fig:CaStats}).  For the repressilator, we actually saw the greatest gains in
situations where stochastic effects were \emph{most\/} prevalent (small $\gamma$\/---Fig.~\ref{fig:RepStatsDet}).  This is a particularly intriguing result.
Gene regulation is a common feature of many biological models and our results indicate a great potential advantage to using leaping in cases of slow
transcription-factor binding and unbinding (such as observed in Ref.~\citep{Blake06}).

A critical aspect of the present study was our ability to identify conditions under which the leaping algorithm did \emph{not\/} perform particularly well.  In
many ways, this may be more valuable in terms of advancing the use of leaping methods than is highlighting its strengths. The leaping algorithm clearly falters
when applied to the full repressilator model (Table~\ref{table:RepRxns}) with large telegraph factor $\gamma$\/ (Fig.~\ref{fig:RepTimings}). Intuitively, it is
easy to understand why this is.  The basic strategy underlying all leaping algorithms is to allow, at each simulation step, as many reaction firings as
possible without the reaction rates in the system changing ``appreciably" \citep{Gillesp00, Gillesp01, Gillesp07}. However, in this case there is only a single
copy of each gene. Thus, only a single binding/unbinding event is possible at each simulation step because one firing changes the binding/unbinding rates from
either finite values to zero or vice versa, which is obviously appreciable.  When $\gamma$\/ is small, this is not a problem because the time interval between
successive binding and unbinding events is large enough so that many transcription, translation and degradation reactions can fire.  When $\gamma$\/ is large,
however, this is no longer the case.  The numbers of reaction firings become limited due to the high frequency of binding and unbinding, and in the extreme
limit the effect is such that the performance of the algorithm approaches that of the SSA (i.e., one reaction firing per step---Fig.~\ref{fig:RepTimings}). We
can generalize this observation by saying that small reaction subnetworks (pairs of reversible reactions in this case) that have small populations and large
rate constants are prime bottlenecks for leaping algorithms.

Fortunately, our results also illustrate how one can surmount such problems.  By applying a simple rapid-equilibrium assumption to the first four reactions of
Table~\ref{table:RepRxns}, we were able to recover the behavior of the full model for $\gamma\!\ge\!1$ (Fig.~\ref{fig:RepStatsAdair}) at significantly reduced
computational cost (Fig.~\ref{fig:RepTimings}) \cite{note:AdairTimings}.  This includes accurately capturing stochastic effects associated with finite numbers
of mRNAs and proteins. Interestingly, we also showed that reducing the model has little effect on the performance of the SSA (Fig.~\ref{fig:RepTimings}). Thus,
the chief benefit to using model reduction in this case was \emph{not\/} in reducing the number of reactions that had to be considered, but rather in
increasing the size of the time step that could be traversed at each simulation step.  This is a different perspective on the issue than is usual and strongly
suggests that leaping and model reduction should be viewed, not as alternative approaches to the problem of timescale separation (as is common), but as
\emph{complementary\/}. Integrating leaping with advanced model-reduction schemes (e.g., \citep{Shib03, Bund03, Cao05:slowSSA, Gout05, Samant05, Weinan05,
Salis05:QSSA, Morelli08}) is thus an area of great future interest.  As a final note, we did observe some (small) disagreement between the PLA and the SSA
results (Figs.~\ref{fig:CaStats}\textit{a\/}, \ref{fig:CaStats}\textit{b} and \ref{fig:RepStatsDet}\textit{d\/}) which may be due to the inclusion of the
reduced reaction types. This is an issue that will be investigated further in the future.

So, given the effectiveness of the leaping algorithm as demonstrated in this article, why are practical applications of leaping so scarce in the literature?
The answer is likely multifaceted. First, the approach is relatively new and many researchers may simply be unaware, or only vaguely aware, of its existence.
Second, newer incarnations of the method are becoming increasingly complex, to the point that, even if aware of their existence, a non-expert may be unable to
implement them. Third, it has been our experience that there is a common misperception that stochastic simulation algorithms can only be applied to sets of
\emph{elementary\/} reaction types. Indeed, it is common practice when investigating the stochastic characteristics of an established biochemical model to
first ``deconstruct" it into elementary reaction steps.  While this is not, in fact, strictly necessary, it is possible that attempts to use leaping algorithms
in this way have befallen the bottleneck of fast-reversible reactions illustrated in Fig.~\ref{fig:RepTimings}.

It is our hope that this article alleviates, to some extent, each of these hindrances to the expanded use of leaping algorithms in computational systems
biology.  For we believe that the future of these methods is bright.  Leaping methods represent a small but important piece of the larger puzzle that is
comprehensible and actionable models of complex biochemical processes. Coupled with advanced model-reduction techniques that address the problem of rate
constant disparities, they can provide a sound and practical means by which the problem of timescale separation in biological systems can be overcome. Further,
imbedded into larger modeling and simulation frameworks that include methods for addressing combinatorial complexity \citep{Hlava06}, spatial localization
\citep{Lemerle05} and parameter uncertainty \citep{Brown03, Guna05}, the promises of \textit{in~silico\/} biology \citep{Palsson00} might finally be within
reach.

\appendix
\section{The Adair Reduction} \label{appx:Adair}

For large telegraph factor $\gamma$\/, the gene-protein binding and unbinding reactions
\begin{eqnarray}
    g + r & \xrightleftharpoons[\gamma k_{-1}]{\gamma k_1} & \{g{\cdot}r\}, \label{rxn:bind1}  \\
    \{g{\cdot}r\} + r & \xrightleftharpoons[\gamma k_{-2}]{\gamma k_2} &
    \{g{\cdot}r{\cdot}r\}, \label{rxn:bind2}
\end{eqnarray}
from Table~\ref{table:RepRxns} can be assumed to be in rapid equilibrium. [Here, we use simpler notation for convenience: $g$\/ for the gene promoter, $r$\/
for the repressor protein and $m$\/ for mRNA (below).]  Setting the forward and reverse rates of reaction pairs~\ref{rxn:bind1} and \ref{rxn:bind2} equal to
each other, it is easy to show that
\begin{equation}
    [g] = K_1 K_2 [g{\cdot}r{\cdot}r]/[r]^2,
\end{equation}
where $K_i\!\equiv\!k_{-i}/k_i$\/ and $[\cdot]$ denotes concentration (or, more correctly, \emph{occupancy probability\/}).  Assuming that the total number of
genes, $g_T$\/, is constant,
\begin{equation}
    g_T = [g] + [g{\cdot}r] + [g{\cdot}r{\cdot}r],
\end{equation}
simple algebra leads to
\begin{equation}
    [g] = \frac{K_1 K_2 g_T}{K_1 K_2 + K_2 [r] + [r]^2}.
    \label{eq:g}
\end{equation}
It is then straightforward to show that
\begin{eqnarray}
    [g{\cdot}r] & = & [g]\frac{[r]}{K_1} = \frac{K_2
    [r] g_T}{K_1 K_2 + K_2 [r] + [r]^2}, \label{eq:gp} \\
    {[g{\cdot}r{\cdot}r]} & = & [g{\cdot}r]\frac{[r]}{K_2}
    = \frac{[r]^2 g_T}{K_1 K_2 + K_2 [r] + [r]^2}. \label{eq:gpp}
\end{eqnarray}
The mRNA transcription reactions are
\begin{eqnarray}
    g & \stackrel{k_\mathrm{tr}}{\longrightarrow} & g + m, \label{rxn:g-mRNA} \\
    \{g{\cdot}r\} & \stackrel{k_\mathrm{tr}^\prime}{\longrightarrow} &
    \{g{\cdot}r\} + m, \\
    \{g{\cdot}r{\cdot}r\} & \stackrel{k_\mathrm{tr}^{\prime\prime}}{\longrightarrow}
    & \{g{\cdot}r{\cdot}r\} + m. \label{rxn:gpp-mRNA}
\end{eqnarray}
The effective rate expressions for mRNA production given in Table~\ref{table:RepReduced} are obtained by multiplying the rate constants in
reactions~\ref{rxn:g-mRNA}--\ref{rxn:gpp-mRNA} by the expressions in Eqs.~\ref{eq:g}--\ref{eq:gpp}.  These effective expressions are of the \emph{Adair\/} form
\citep{Bowden04}.

Notice that Eq.~\ref{eq:gpp} is similar in form to a Hill equation \citep{Bowden04} with a Hill coefficient of +2 (positive cooperativity), except for the $K_2
[r]$ term in the denominator. However, this term vanishes as $[r]\!\rightarrow\!0$, dominated by $K_1 K_2$, and is overwhelmed by $[r]^2$ as
$[r]\!\rightarrow\!\infty$. Thus, this term can effectively be ignored in all cases other than a small range of intermediate levels of $[r]$. This can be seen,
therefore, as a type of ``derivation" of the Hill equation. Similarly, Eq.~\ref{eq:g} can be rewritten as
\begin{equation}
    [g] = \frac{[r]^{-2} g_T}{[r]^{-2} + (\sqrt{K_1 [r]})^{-2}
        + (\sqrt{K_1 K_2})^{-2}}, \label{eq:g-Hillform}
\end{equation}
which is similar to a negative-cooperativity Hill equation with a Hill coefficient of $-2$, again except for the second term in the denominator.  Note that the
expression for $[g{\cdot}r]$ in Eq.~\ref{eq:gp} is intermediate between Eqs.~\ref{eq:gpp} and \ref{eq:g-Hillform} and does not have an analogous Hill form.

\section{Michaelis-Menten, Adair, and $\tau$\/~Selection} \label{appx:tauSelec}

At each step of a PLA simulation, a time step $\tau$\/ is calculated based on the current state of the system.  This time step corresponds to the time interval
over which we expect the reaction rates $a_\mu$\/ (``propensities" in the stochastic jargon) for all reactions in a system to remain \emph{essentially\/}
constant \citep{Gillesp01, Gillesp03, Cao06:newStep, Harris06}.  We quantify the concept of ``essentially constant" by imposing a constraint on the relative
change of each reaction propensity \citep{Cao06:newStep},
\begin{equation}
    \left|\,a_\mu(t+\tau)-a_\mu(t)\,\right| / a_\mu(t) =
    \epsilon \,\,\,\,\,\, (0 < \epsilon \ll 1).
    \label{eq:RBconstraint}
\end{equation}
In practice, there are two approaches for determining $\tau$\/. The first is a ``reaction-based" approach in which the constraint in Eq.~\ref{eq:RBconstraint}
is used directly \citep{Gillesp01, Gillesp03, Cao06:newStep}.  The second, which is used in this article, is a ``species-based" approach where the relative
changes in each reactant population $X_i$\/ are constrained such that Eq.~\ref{eq:RBconstraint} is satisfied for all reactions \citep{Cao06:newStep},
\begin{equation}
    \left|\,X_i(t+\tau)-X_i(t)\,\right| / X_i(t) =
    \epsilon/g_i.
    \label{eq:SBconstraint}
\end{equation}
Here, $\epsilon$\/ is the same as in Eq.~\ref{eq:RBconstraint} and $g_i$\/ depends on the highest-order reaction species $S_i$\/ is involved in.

Procedures for determining $g_i$\/ which account for elementary reaction types up to third order are given in \citet{Cao06:newStep} and, in modified form, in
\citet{Harris06}. Basically, each reaction type has associated with it a value of $g_i$\/ for each reactant species $S_i$\/.  Thus, one merely sifts through
all reactions in which $S_i$\/ appears as a reactant and sets $g_i$\/ equal to the largest of these values. This need be done only once, at the outset of a
simulation. In this article, we consider non-elementary reactions of the Michaelis-Menten (MM) and Adair types (Tables~\ref{table:CaRxns} and
\ref{table:RepReduced}) and must, therefore, derive appropriate $g_i$\/ expressions for them.

In Table~\ref{table:CaRxns}, we consider two different types of Michaelis-Menten reactions, which we can term 1st-order (e.g.,
$\mathrm{PLC}^*\!\rightarrow\!\emptyset$) and 2nd-order (e.g., $G_\alpha\!+\!\mathrm{PLC}^*\!\rightarrow\!\mathrm{PLC}^*$) MM types.  It is easy to show that
$g_i$\/ for each of these is the same as for the corresponding elementary reaction \citep{Cao06:newStep, Harris06}, i.e., $g_i\!=\!1$ for 1st-order MM and
$g_i\!=\!2$ for 2nd-order MM. For 1st-order MM reactions we have
\begin{eqnarray}
    a_\mu & = & \left( \frac{c_\mathrm{cat}}{X_i+C_M} \right) X_i, \nonumber \\
    \Delta a_\mu & = & \frac{da_\mu}{dX_i} \Delta X_i
        = a_\mu \left( \frac{\Delta X_i}{X_i} - \frac{\Delta
          X_i}{X_i+C_M} \right), \nonumber \\
    \frac{|\Delta a_\mu|}{a_\mu} & = & \left( \frac{C_M}{X_i+C_M}
        \right) \frac{|\Delta X_i|}{X_i} \leq \frac{|\Delta X_i|}{X_i}.
        \label{eq:gi-1stMM}
\end{eqnarray}
Equation~\ref{eq:gi-1stMM} shows that if we constrain $|\Delta X_i|/X_i\!=\!\epsilon$\/, then $|\Delta a_\mu|/a_\mu\!\leq\!\epsilon$\/, i.e., $g_i\!=\!1$ will
suffice. Similarly, for 2nd-order MM reactions,
\begin{eqnarray}
    a_\mu & = & \left( \frac{c_\mathrm{cat}}{X_i+C_M} \right) X_i X_j, \nonumber \\
    \Delta a_\mu & \approx & \frac{\partial a_\mu}{\partial X_i} \Delta X_i
        + \frac{\partial a_\mu}{\partial X_j} \Delta X_j, \nonumber \\
    \frac{|\Delta a_\mu|}{a_\mu} & \lesssim & \left( \frac{C_M}{X_i+C_M}
        \right) \frac{|\Delta X_i|}{X_i} + \frac{|\Delta X_j|}{X_j}.
        \label{eq:gi-2ndMM}
\end{eqnarray}
Thus, if we constrain both $|\Delta X_i|/X_i\!=\!|\Delta X_j|/X_j\!=\!\epsilon/2$, then $|\Delta a_\mu|/a_\mu\!\leq\!\epsilon$\/, i.e., $g_i\!=\!g_j\!=\!2$.

The Adair reactions in Table~\ref{table:RepReduced} represent effective rates of mRNA production from the unbound, singly-bound and doubly-bound gene.  In
general terms, we can think of the rates of mRNA production from an $n$\/-bound gene with $m$\/ binding sites ($0\!\leq\!n\!\leq\!m$). We then have
\begin{equation}
    a_\mu^{n,m}(r) = \frac{ r^n g_T c_\mathrm{cat} \prod_{i=n}^{m-1} C_i }
                    { \sum_{i=0}^m \left( r^i \prod_{j=i}^{m-1} C_j \right)},
                    \label{eq:a-Adair}
\end{equation}
where $r$\/ is the repressor protein population, $g_T$\/ is the total number of genes (unity in this case), and $C_i\!=\!K_i\!\times\!N_A\Omega$.  It is easy
to show that for $m\!=\!2$ and $n\!=\!\{0,1,2\}$, Eq.~\ref{eq:a-Adair} reduces to the expressions given in Table~\ref{table:RepReduced}.

Following the same procedure as above, we get
\begin{align}
    \frac{|\Delta a_\mu^{n,m}|}{a_\mu^{n,m}} & =
    \left|\frac{da_\mu^{n,m}}{dr} \frac{\Delta r}{a_\mu^{n,m}}\right| \nonumber \\
    & = \left| n - \frac{m \sum_{i=1}^m \left( \frac{i}{m} r^i
    \prod_{j=i}^{m-1} C_j \right)}{\prod_{j=0}^{m-1} C_j +
    \sum_{i=1}^m \left( r^i \prod_{j=i}^{m-1} C_j \right)} \right|
    \frac{|\Delta r|}{r}. \label{eq:dadp-Adair}
\end{align}
From Eq.~\ref{eq:dadp-Adair}, we see that as $r\!\rightarrow\!0$, $\frac{|\Delta a_\mu^{n,m}|}{a_\mu^{n,m}} \rightarrow n\frac{|\Delta r|}{r},$ and as
$r\!\rightarrow\!\infty$, $\frac{|\Delta a_\mu^{n,m}|}{a_\mu^{n,m}} \rightarrow |n-m|\frac{|\Delta r|}{r}$.  This means that if we constrain $|\Delta
r|/r\!=\!\epsilon/\mathrm{max}\{n,|n-m|\}$ then $|\Delta a_\mu|/a_\mu\!\leq\!\epsilon$\/ in all cases, i.e., $g_i\!=\!\mathrm{max}\{n,|n-m|\}$.  For the three
Adair reactions in Table~\ref{table:RepReduced}, this gives $g_i\!=\!\{2,1,2\}$ for $n\!=\!\{0,1,2\}$, respectively.

\begin{acknowledgments}
We thank H.~Lee, J.~E.\ Goose, K.~A.\ Iyengar, F.~P.\ Casey and J.~P.\ Sethna for useful discussions regarding this work and acknowledge financial support from
the Semiconductor Research Corporation.  L.A.H.\ extends special thanks to Prof.\ J.~R.\ Faeder.  A.M.P.\ and E.R.M.\ further thank the Intel Corporation for
funding through Cornell's Learning Initiatives for Future Engineers (LIFE) program.
\end{acknowledgments}



\begin{thebibliography}{93}
\expandafter\ifx\csname natexlab\endcsname\relax\def\natexlab#1{#1}\fi \expandafter\ifx\csname bibnamefont\endcsname\relax
  \def\bibnamefont#1{#1}\fi
\expandafter\ifx\csname bibfnamefont\endcsname\relax
  \def\bibfnamefont#1{#1}\fi
\expandafter\ifx\csname citenamefont\endcsname\relax
  \def\citenamefont#1{#1}\fi
\expandafter\ifx\csname url\endcsname\relax
  \def\url#1{\texttt{#1}}\fi
\expandafter\ifx\csname urlprefix\endcsname\relax\def\urlprefix{URL }\fi \providecommand{\bibinfo}[2]{#2} \providecommand{\eprint}[2][]{\url{#2}}

\bibitem[{\citenamefont{McAdams and Arkin}(1999)}]{McAdams99}
\bibinfo{author}{\bibfnamefont{H.~H.} \bibnamefont{McAdams}} \bibnamefont{and}
  \bibinfo{author}{\bibfnamefont{A.}~\bibnamefont{Arkin}},
  \bibinfo{journal}{Trends Genet.} \textbf{\bibinfo{volume}{15}},
  \bibinfo{pages}{65} (\bibinfo{year}{1999}).

\bibitem[{\citenamefont{Rao et~al.}(2002)\citenamefont{Rao, Wolf, and
  Arkin}}]{Rao02}
\bibinfo{author}{\bibfnamefont{C.~V.} \bibnamefont{Rao}},
  \bibinfo{author}{\bibfnamefont{D.~M.} \bibnamefont{Wolf}}, \bibnamefont{and}
  \bibinfo{author}{\bibfnamefont{A.~P.} \bibnamefont{Arkin}},
  \bibinfo{journal}{Nature} \textbf{\bibinfo{volume}{420}},
  \bibinfo{pages}{231} (\bibinfo{year}{2002}).

\bibitem[{\citenamefont{Raser and O'Shea}(2005)}]{Raser05}
\bibinfo{author}{\bibfnamefont{J.~M.} \bibnamefont{Raser}} \bibnamefont{and}
  \bibinfo{author}{\bibfnamefont{E.~K.} \bibnamefont{O'Shea}},
  \bibinfo{journal}{Science} \textbf{\bibinfo{volume}{309}},
  \bibinfo{pages}{2010} (\bibinfo{year}{2005}).

\bibitem[{\citenamefont{K{\ae}rn et~al.}(2005)\citenamefont{K{\ae}rn, Elston,
  Blake, and Collins}}]{Kaern05}
\bibinfo{author}{\bibfnamefont{M.}~\bibnamefont{K{\ae}rn}},
  \bibinfo{author}{\bibfnamefont{T.~C.} \bibnamefont{Elston}},
  \bibinfo{author}{\bibfnamefont{W.~J.} \bibnamefont{Blake}}, \bibnamefont{and}
  \bibinfo{author}{\bibfnamefont{J.~J.} \bibnamefont{Collins}},
  \bibinfo{journal}{Nature Rev. Genet.} \textbf{\bibinfo{volume}{6}},
  \bibinfo{pages}{451} (\bibinfo{year}{2005}).

\bibitem[{\citenamefont{{Di~Ventura} et~al.}(2006)\citenamefont{{Di~Ventura},
  Lemerle, Michalodimitrakis, and Serrano}}]{Vent06}
\bibinfo{author}{\bibfnamefont{B.}~\bibnamefont{{Di~Ventura}}},
  \bibinfo{author}{\bibfnamefont{C.}~\bibnamefont{Lemerle}},
  \bibinfo{author}{\bibfnamefont{K.}~\bibnamefont{Michalodimitrakis}},
  \bibnamefont{and} \bibinfo{author}{\bibfnamefont{L.}~\bibnamefont{Serrano}},
  \bibinfo{journal}{Nature} \textbf{\bibinfo{volume}{443}},
  \bibinfo{pages}{527} (\bibinfo{year}{2006}).

\bibitem[{\citenamefont{Samoilov and Arkin}(2006)}]{Samo06:Dev}
\bibinfo{author}{\bibfnamefont{M.~S.} \bibnamefont{Samoilov}} \bibnamefont{and}
  \bibinfo{author}{\bibfnamefont{A.~P.} \bibnamefont{Arkin}},
  \bibinfo{journal}{Nature Biotechnol.} \textbf{\bibinfo{volume}{24}},
  \bibinfo{pages}{1235} (\bibinfo{year}{2006}).

\bibitem[{\citenamefont{Samoilov et~al.}(2006)\citenamefont{Samoilov, Price,
  and Arkin}}]{Samo06:STKE}
\bibinfo{author}{\bibfnamefont{M.~S.} \bibnamefont{Samoilov}},
  \bibinfo{author}{\bibfnamefont{G.}~\bibnamefont{Price}}, \bibnamefont{and}
  \bibinfo{author}{\bibfnamefont{A.~P.} \bibnamefont{Arkin}},
  \bibinfo{journal}{Sci. STKE} \textbf{\bibinfo{volume}{2006
  \textnormal{(366)}}}, \bibinfo{pages}{re17} (\bibinfo{year}{2006}).

\bibitem[{\citenamefont{Maheshri and O'Shea}(2007)}]{Mahesh07}
\bibinfo{author}{\bibfnamefont{N.}~\bibnamefont{Maheshri}} \bibnamefont{and}
  \bibinfo{author}{\bibfnamefont{E.~K.} \bibnamefont{O'Shea}},
  \bibinfo{journal}{Annu. Rev. Biophys. Biomol. Struct.}
  \textbf{\bibinfo{volume}{36}}, \bibinfo{pages}{413} (\bibinfo{year}{2007}).

\bibitem[{\citenamefont{Spudich and {Koshland,~Jr.}}(1976)}]{Spud76}
\bibinfo{author}{\bibfnamefont{J.~L.} \bibnamefont{Spudich}} \bibnamefont{and}
  \bibinfo{author}{\bibfnamefont{D.~E.} \bibnamefont{{Koshland,~Jr.}}},
  \bibinfo{journal}{Nature} \textbf{\bibinfo{volume}{262}},
  \bibinfo{pages}{467} (\bibinfo{year}{1976}).

\bibitem[{\citenamefont{Elowitz et~al.}(2002)\citenamefont{Elowitz, Levine,
  Siggia, and Swain}}]{Elow02}
\bibinfo{author}{\bibfnamefont{M.~B.} \bibnamefont{Elowitz}},
  \bibinfo{author}{\bibfnamefont{A.~J.} \bibnamefont{Levine}},
  \bibinfo{author}{\bibfnamefont{E.~D.} \bibnamefont{Siggia}},
  \bibnamefont{and} \bibinfo{author}{\bibfnamefont{P.~S.} \bibnamefont{Swain}},
  \bibinfo{journal}{Science} \textbf{\bibinfo{volume}{297}},
  \bibinfo{pages}{1183} (\bibinfo{year}{2002}).

\bibitem[{\citenamefont{Fedoroff and Fontana}(2002)}]{Fedo02}
\bibinfo{author}{\bibfnamefont{N.}~\bibnamefont{Fedoroff}} \bibnamefont{and}
  \bibinfo{author}{\bibfnamefont{W.}~\bibnamefont{Fontana}},
  \bibinfo{journal}{Science} \textbf{\bibinfo{volume}{297}},
  \bibinfo{pages}{1129} (\bibinfo{year}{2002}).

\bibitem[{\citenamefont{Blake et~al.}(2006)\citenamefont{Blake, Bal\'{a}zsi,
  Kohanski, Isaacs, Murphy, Kuang, Cantor, Walt, and Collins}}]{Blake06}
\bibinfo{author}{\bibfnamefont{W.~J.} \bibnamefont{Blake}},
  \bibinfo{author}{\bibfnamefont{G.}~\bibnamefont{Bal\'{a}zsi}},
  \bibinfo{author}{\bibfnamefont{M.~A.} \bibnamefont{Kohanski}},
  \bibinfo{author}{\bibfnamefont{F.~J.} \bibnamefont{Isaacs}},
  \bibinfo{author}{\bibfnamefont{K.~F.} \bibnamefont{Murphy}},
  \bibinfo{author}{\bibfnamefont{Y.}~\bibnamefont{Kuang}},
  \bibinfo{author}{\bibfnamefont{C.~R.} \bibnamefont{Cantor}},
  \bibinfo{author}{\bibfnamefont{D.~R.} \bibnamefont{Walt}}, \bibnamefont{and}
  \bibinfo{author}{\bibfnamefont{J.~J.} \bibnamefont{Collins}},
  \bibinfo{journal}{Mol. Cell} \textbf{\bibinfo{volume}{24}},
  \bibinfo{pages}{853} (\bibinfo{year}{2006}).

\bibitem[{\citenamefont{Elowitz and Leibler}(2000)}]{Elow00}
\bibinfo{author}{\bibfnamefont{M.~B.} \bibnamefont{Elowitz}} \bibnamefont{and}
  \bibinfo{author}{\bibfnamefont{S.}~\bibnamefont{Leibler}},
  \bibinfo{journal}{Nature} \textbf{\bibinfo{volume}{403}},
  \bibinfo{pages}{335} (\bibinfo{year}{2000}).

\bibitem[{\citenamefont{Barkai and Leibler}(2000)}]{Barkai00}
\bibinfo{author}{\bibfnamefont{N.}~\bibnamefont{Barkai}} \bibnamefont{and}
  \bibinfo{author}{\bibfnamefont{S.}~\bibnamefont{Leibler}},
  \bibinfo{journal}{Nature} \textbf{\bibinfo{volume}{403}},
  \bibinfo{pages}{267} (\bibinfo{year}{2000}).

\bibitem[{\citenamefont{Vilar et~al.}(2002)\citenamefont{Vilar, Kueh, Barkai,
  and Leibler}}]{Vilar02}
\bibinfo{author}{\bibfnamefont{J.~M.~G.} \bibnamefont{Vilar}},
  \bibinfo{author}{\bibfnamefont{H.~Y.} \bibnamefont{Kueh}},
  \bibinfo{author}{\bibfnamefont{N.}~\bibnamefont{Barkai}}, \bibnamefont{and}
  \bibinfo{author}{\bibfnamefont{S.}~\bibnamefont{Leibler}},
  \bibinfo{journal}{Proc. Natl. Acad. Sci. USA} \textbf{\bibinfo{volume}{99}},
  \bibinfo{pages}{5988} (\bibinfo{year}{2002}).

\bibitem[{\citenamefont{Gonze et~al.}(2002)\citenamefont{Gonze, Halloy, and
  Goldbeter}}]{Gonze02:PNAS}
\bibinfo{author}{\bibfnamefont{D.}~\bibnamefont{Gonze}},
  \bibinfo{author}{\bibfnamefont{J.}~\bibnamefont{Halloy}}, \bibnamefont{and}
  \bibinfo{author}{\bibfnamefont{A.}~\bibnamefont{Goldbeter}},
  \bibinfo{journal}{Proc. Natl. Acad. Sci. USA} \textbf{\bibinfo{volume}{99}},
  \bibinfo{pages}{673} (\bibinfo{year}{2002}).

\bibitem[{\citenamefont{Aldridge et~al.}(2006)\citenamefont{Aldridge, Burke,
  Lauffenburger, and Sorger}}]{Aldr06}
\bibinfo{author}{\bibfnamefont{B.~B.} \bibnamefont{Aldridge}},
  \bibinfo{author}{\bibfnamefont{J.~M.} \bibnamefont{Burke}},
  \bibinfo{author}{\bibfnamefont{D.~A.} \bibnamefont{Lauffenburger}},
  \bibnamefont{and} \bibinfo{author}{\bibfnamefont{P.~K.}
  \bibnamefont{Sorger}}, \bibinfo{journal}{Nature Cell Biol.}
  \textbf{\bibinfo{volume}{8}}, \bibinfo{pages}{1195} (\bibinfo{year}{2006}).

\bibitem[{\citenamefont{Tomlin and Axelrod}(2007)}]{Tomlin07}
\bibinfo{author}{\bibfnamefont{C.~J.} \bibnamefont{Tomlin}} \bibnamefont{and}
  \bibinfo{author}{\bibfnamefont{J.~D.} \bibnamefont{Axelrod}},
  \bibinfo{journal}{Nature Rev. Genet.} \textbf{\bibinfo{volume}{8}},
  \bibinfo{pages}{331} (\bibinfo{year}{2007}).

\bibitem[{\citenamefont{Gillespie}(1976)}]{Gillesp76}
\bibinfo{author}{\bibfnamefont{D.~T.} \bibnamefont{Gillespie}},
  \bibinfo{journal}{J. Comput. Phys.} \textbf{\bibinfo{volume}{22}},
  \bibinfo{pages}{403} (\bibinfo{year}{1976}).

\bibitem[{\citenamefont{Gillespie}(1977)}]{Gillesp77}
\bibinfo{author}{\bibfnamefont{D.~T.} \bibnamefont{Gillespie}},
  \bibinfo{journal}{J. Phys. Chem.} \textbf{\bibinfo{volume}{81}},
  \bibinfo{pages}{2340} (\bibinfo{year}{1977}).

\bibitem[{\citenamefont{Gillespie}(2007)}]{Gillesp07}
\bibinfo{author}{\bibfnamefont{D.~T.} \bibnamefont{Gillespie}},
  \bibinfo{journal}{Annu. Rev. Phys. Chem.} \textbf{\bibinfo{volume}{58}},
  \bibinfo{pages}{35} (\bibinfo{year}{2007}).

\bibitem[{\citenamefont{Endy and Brent}(2001)}]{Endy01}
\bibinfo{author}{\bibfnamefont{D.}~\bibnamefont{Endy}} \bibnamefont{and}
  \bibinfo{author}{\bibfnamefont{R.}~\bibnamefont{Brent}},
  \bibinfo{journal}{Nature} \textbf{\bibinfo{volume}{409}},
  \bibinfo{pages}{391} (\bibinfo{year}{2001}).

\bibitem[{\citenamefont{Gillespie}(2001)}]{Gillesp01}
\bibinfo{author}{\bibfnamefont{D.~T.} \bibnamefont{Gillespie}},
  \bibinfo{journal}{J. Chem. Phys.} \textbf{\bibinfo{volume}{115}},
  \bibinfo{pages}{1716} (\bibinfo{year}{2001}).

\bibitem[{\citenamefont{Gibson and Bruck}(2000)}]{Gibson00}
\bibinfo{author}{\bibfnamefont{M.~A.} \bibnamefont{Gibson}} \bibnamefont{and}
  \bibinfo{author}{\bibfnamefont{J.}~\bibnamefont{Bruck}}, \bibinfo{journal}{J.
  Phys. Chem. A} \textbf{\bibinfo{volume}{104}}, \bibinfo{pages}{1876}
  (\bibinfo{year}{2000}).

\bibitem[{\citenamefont{Resat et~al.}(2001)\citenamefont{Resat, Wiley, and
  Dixon}}]{Resat01}
\bibinfo{author}{\bibfnamefont{H.}~\bibnamefont{Resat}},
  \bibinfo{author}{\bibfnamefont{H.~S.} \bibnamefont{Wiley}}, \bibnamefont{and}
  \bibinfo{author}{\bibfnamefont{D.~A.} \bibnamefont{Dixon}},
  \bibinfo{journal}{J. Phys. Chem. B} \textbf{\bibinfo{volume}{105}},
  \bibinfo{pages}{11026} (\bibinfo{year}{2001}).

\bibitem[{\citenamefont{Cao et~al.}(2004{\natexlab{a}})\citenamefont{Cao, Li,
  and Petzold}}]{Cao04:ODM}
\bibinfo{author}{\bibfnamefont{Y.}~\bibnamefont{Cao}},
  \bibinfo{author}{\bibfnamefont{H.}~\bibnamefont{Li}}, \bibnamefont{and}
  \bibinfo{author}{\bibfnamefont{L.}~\bibnamefont{Petzold}},
  \bibinfo{journal}{J. Chem. Phys.} \textbf{\bibinfo{volume}{121}},
  \bibinfo{pages}{4059} (\bibinfo{year}{2004}{\natexlab{a}}).

\bibitem[{\citenamefont{McCollum et~al.}(2006)\citenamefont{McCollum, Peterson,
  Cox, Simpson, and Samatova}}]{McColl06}
\bibinfo{author}{\bibfnamefont{J.~M.} \bibnamefont{McCollum}},
  \bibinfo{author}{\bibfnamefont{G.~D.} \bibnamefont{Peterson}},
  \bibinfo{author}{\bibfnamefont{C.~D.} \bibnamefont{Cox}},
  \bibinfo{author}{\bibfnamefont{M.~L.} \bibnamefont{Simpson}},
  \bibnamefont{and} \bibinfo{author}{\bibfnamefont{N.~F.}
  \bibnamefont{Samatova}}, \bibinfo{journal}{Comput. Biol. Chem.}
  \textbf{\bibinfo{volume}{30}}, \bibinfo{pages}{39} (\bibinfo{year}{2006}).

\bibitem[{\citenamefont{Gillespie and Petzold}(2003)}]{Gillesp03}
\bibinfo{author}{\bibfnamefont{D.~T.} \bibnamefont{Gillespie}}
  \bibnamefont{and} \bibinfo{author}{\bibfnamefont{L.~R.}
  \bibnamefont{Petzold}}, \bibinfo{journal}{J. Chem. Phys.}
  \textbf{\bibinfo{volume}{119}}, \bibinfo{pages}{8229} (\bibinfo{year}{2003}).

\bibitem[{\citenamefont{Rathinam et~al.}(2003)\citenamefont{Rathinam, Petzold,
  Cao, and Gillespie}}]{Rath03}
\bibinfo{author}{\bibfnamefont{M.}~\bibnamefont{Rathinam}},
  \bibinfo{author}{\bibfnamefont{L.~R.} \bibnamefont{Petzold}},
  \bibinfo{author}{\bibfnamefont{Y.}~\bibnamefont{Cao}}, \bibnamefont{and}
  \bibinfo{author}{\bibfnamefont{D.~T.} \bibnamefont{Gillespie}},
  \bibinfo{journal}{J. Chem. Phys.} \textbf{\bibinfo{volume}{119}},
  \bibinfo{pages}{12784} (\bibinfo{year}{2003}).

\bibitem[{\citenamefont{Cao et~al.}(2004{\natexlab{b}})\citenamefont{Cao,
  Petzold, Rathinam, and Gillespie}}]{Cao04:Stability}
\bibinfo{author}{\bibfnamefont{Y.}~\bibnamefont{Cao}},
  \bibinfo{author}{\bibfnamefont{L.~R.} \bibnamefont{Petzold}},
  \bibinfo{author}{\bibfnamefont{M.}~\bibnamefont{Rathinam}}, \bibnamefont{and}
  \bibinfo{author}{\bibfnamefont{D.~T.} \bibnamefont{Gillespie}},
  \bibinfo{journal}{J. Chem. Phys.} \textbf{\bibinfo{volume}{121}},
  \bibinfo{pages}{12169} (\bibinfo{year}{2004}{\natexlab{b}}).

\bibitem[{\citenamefont{Tian and Burrage}(2004)}]{Tian04}
\bibinfo{author}{\bibfnamefont{T.}~\bibnamefont{Tian}} \bibnamefont{and}
  \bibinfo{author}{\bibfnamefont{K.}~\bibnamefont{Burrage}},
  \bibinfo{journal}{J. Chem. Phys.} \textbf{\bibinfo{volume}{121}},
  \bibinfo{pages}{10356} (\bibinfo{year}{2004}).

\bibitem[{\citenamefont{Chatterjee
  et~al.}(2005{\natexlab{a}})\citenamefont{Chatterjee, Vlachos, and
  Katsoulakis}}]{Chatt05}
\bibinfo{author}{\bibfnamefont{A.}~\bibnamefont{Chatterjee}},
  \bibinfo{author}{\bibfnamefont{D.~G.} \bibnamefont{Vlachos}},
  \bibnamefont{and} \bibinfo{author}{\bibfnamefont{M.~A.}
  \bibnamefont{Katsoulakis}}, \bibinfo{journal}{J. Chem. Phys.}
  \textbf{\bibinfo{volume}{122}}, \bibinfo{pages}{024112}
  (\bibinfo{year}{2005}{\natexlab{a}}).

\bibitem[{\citenamefont{Cao et~al.}(2005{\natexlab{a}})\citenamefont{Cao,
  Gillespie, and Petzold}}]{Cao05:negPop}
\bibinfo{author}{\bibfnamefont{Y.}~\bibnamefont{Cao}},
  \bibinfo{author}{\bibfnamefont{D.~T.} \bibnamefont{Gillespie}},
  \bibnamefont{and} \bibinfo{author}{\bibfnamefont{L.~R.}
  \bibnamefont{Petzold}}, \bibinfo{journal}{J. Chem. Phys.}
  \textbf{\bibinfo{volume}{123}}, \bibinfo{pages}{054104}
  (\bibinfo{year}{2005}{\natexlab{a}}).

\bibitem[{\citenamefont{Cao et~al.}(2006)\citenamefont{Cao, Gillespie, and
  Petzold}}]{Cao06:newStep}
\bibinfo{author}{\bibfnamefont{Y.}~\bibnamefont{Cao}},
  \bibinfo{author}{\bibfnamefont{D.~T.} \bibnamefont{Gillespie}},
  \bibnamefont{and} \bibinfo{author}{\bibfnamefont{L.~R.}
  \bibnamefont{Petzold}}, \bibinfo{journal}{J. Chem. Phys.}
  \textbf{\bibinfo{volume}{124}}, \bibinfo{pages}{044109}
  (\bibinfo{year}{2006}).

\bibitem[{\citenamefont{Wagner et~al.}(2006)\citenamefont{Wagner, M\"{o}ller,
  and Prank}}]{Wagner06}
\bibinfo{author}{\bibfnamefont{H.}~\bibnamefont{Wagner}},
  \bibinfo{author}{\bibfnamefont{M.}~\bibnamefont{M\"{o}ller}},
  \bibnamefont{and} \bibinfo{author}{\bibfnamefont{K.}~\bibnamefont{Prank}},
  \bibinfo{journal}{J. Chem. Phys.} \textbf{\bibinfo{volume}{125}},
  \bibinfo{pages}{174104} (\bibinfo{year}{2006}).

\bibitem[{\citenamefont{Auger et~al.}(2006)\citenamefont{Auger, Chatelain, and
  Koumoutsakos}}]{Auger06}
\bibinfo{author}{\bibfnamefont{A.}~\bibnamefont{Auger}},
  \bibinfo{author}{\bibfnamefont{P.}~\bibnamefont{Chatelain}},
  \bibnamefont{and}
  \bibinfo{author}{\bibfnamefont{P.}~\bibnamefont{Koumoutsakos}},
  \bibinfo{journal}{J. Chem. Phys.} \textbf{\bibinfo{volume}{125}},
  \bibinfo{pages}{084103} (\bibinfo{year}{2006}).

\bibitem[{\citenamefont{Cai and Xu}(2007)}]{Cai07}
\bibinfo{author}{\bibfnamefont{X.}~\bibnamefont{Cai}} \bibnamefont{and}
  \bibinfo{author}{\bibfnamefont{Z.}~\bibnamefont{Xu}}, \bibinfo{journal}{J.
  Chem. Phys.} \textbf{\bibinfo{volume}{126}}, \bibinfo{pages}{074102}
  (\bibinfo{year}{2007}).

\bibitem[{\citenamefont{Pettigrew and Resat}(2007)}]{Petti07}
\bibinfo{author}{\bibfnamefont{M.~F.} \bibnamefont{Pettigrew}}
  \bibnamefont{and} \bibinfo{author}{\bibfnamefont{H.}~\bibnamefont{Resat}},
  \bibinfo{journal}{J. Chem. Phys.} \textbf{\bibinfo{volume}{126}},
  \bibinfo{pages}{084101} (\bibinfo{year}{2007}).

\bibitem[{\citenamefont{Peng et~al.}(2007)\citenamefont{Peng, Zhou, and
  Wang}}]{Peng07}
\bibinfo{author}{\bibfnamefont{X.}~\bibnamefont{Peng}},
  \bibinfo{author}{\bibfnamefont{W.}~\bibnamefont{Zhou}}, \bibnamefont{and}
  \bibinfo{author}{\bibfnamefont{Y.}~\bibnamefont{Wang}}, \bibinfo{journal}{J.
  Chem. Phys.} \textbf{\bibinfo{volume}{126}}, \bibinfo{pages}{224109}
  (\bibinfo{year}{2007}).

\bibitem[{\citenamefont{Cao et~al.}(2007)\citenamefont{Cao, Gillespie, and
  Petzold}}]{Cao07}
\bibinfo{author}{\bibfnamefont{Y.}~\bibnamefont{Cao}},
  \bibinfo{author}{\bibfnamefont{D.~T.} \bibnamefont{Gillespie}},
  \bibnamefont{and} \bibinfo{author}{\bibfnamefont{L.~R.}
  \bibnamefont{Petzold}}, \bibinfo{journal}{J. Chem. Phys.}
  \textbf{\bibinfo{volume}{126}}, \bibinfo{pages}{224101}
  (\bibinfo{year}{2007}).

\bibitem[{\citenamefont{Rathinam and {El~Samad}}(2007)}]{Rath07}
\bibinfo{author}{\bibfnamefont{M.}~\bibnamefont{Rathinam}} \bibnamefont{and}
  \bibinfo{author}{\bibfnamefont{H.}~\bibnamefont{{El~Samad}}},
  \bibinfo{journal}{J. Comput. Phys.} \textbf{\bibinfo{volume}{224}},
  \bibinfo{pages}{897} (\bibinfo{year}{2007}).

\bibitem[{\citenamefont{Anderson}(2008)}]{Ander08}
\bibinfo{author}{\bibfnamefont{D.~F.} \bibnamefont{Anderson}},
  \bibinfo{journal}{J. Chem. Phys.} \textbf{\bibinfo{volume}{128}},
  \bibinfo{pages}{054103} (\bibinfo{year}{2008}).

\bibitem[{\citenamefont{Xu and Cai}(2008)}]{Xu08}
\bibinfo{author}{\bibfnamefont{Z.}~\bibnamefont{Xu}} \bibnamefont{and}
  \bibinfo{author}{\bibfnamefont{X.}~\bibnamefont{Cai}}, \bibinfo{journal}{J.
  Chem. Phys.} \textbf{\bibinfo{volume}{128}}, \bibinfo{pages}{154112}
  (\bibinfo{year}{2008}).

\bibitem[{\citenamefont{Leier et~al.}(2008)\citenamefont{Leier, {Marquez-Lago},
  and Burrage}}]{Leier08}
\bibinfo{author}{\bibfnamefont{A.}~\bibnamefont{Leier}},
  \bibinfo{author}{\bibfnamefont{T.~T.} \bibnamefont{{Marquez-Lago}}},
  \bibnamefont{and} \bibinfo{author}{\bibfnamefont{K.}~\bibnamefont{Burrage}},
  \bibinfo{journal}{J. Chem. Phys.} \textbf{\bibinfo{volume}{128}},
  \bibinfo{pages}{205107} (\bibinfo{year}{2008}).

\bibitem[{\citenamefont{Harris and Clancy}(2006)}]{Harris06}
\bibinfo{author}{\bibfnamefont{L.~A.} \bibnamefont{Harris}} \bibnamefont{and}
  \bibinfo{author}{\bibfnamefont{P.}~\bibnamefont{Clancy}},
  \bibinfo{journal}{J. Chem. Phys.} \textbf{\bibinfo{volume}{125}},
  \bibinfo{pages}{144107} (\bibinfo{year}{2006}).

\bibitem[{\citenamefont{Haseltine and Rawlings}(2002)}]{Hasel02}
\bibinfo{author}{\bibfnamefont{E.~L.} \bibnamefont{Haseltine}}
  \bibnamefont{and} \bibinfo{author}{\bibfnamefont{J.~B.}
  \bibnamefont{Rawlings}}, \bibinfo{journal}{J. Chem. Phys.}
  \textbf{\bibinfo{volume}{117}}, \bibinfo{pages}{6959} (\bibinfo{year}{2002}).

\bibitem[{\citenamefont{Kiehl et~al.}(2004)\citenamefont{Kiehl, Mattheyses, and
  Simmons}}]{Kiehl04}
\bibinfo{author}{\bibfnamefont{T.~R.} \bibnamefont{Kiehl}},
  \bibinfo{author}{\bibfnamefont{R.~M.} \bibnamefont{Mattheyses}},
  \bibnamefont{and} \bibinfo{author}{\bibfnamefont{M.~K.}
  \bibnamefont{Simmons}}, \bibinfo{journal}{Bioinformatics}
  \textbf{\bibinfo{volume}{20}}, \bibinfo{pages}{316} (\bibinfo{year}{2004}).

\bibitem[{\citenamefont{Takahashi et~al.}(2004)\citenamefont{Takahashi, Kaizu,
  Hu, and Tomita}}]{Taka04}
\bibinfo{author}{\bibfnamefont{K.}~\bibnamefont{Takahashi}},
  \bibinfo{author}{\bibfnamefont{K.}~\bibnamefont{Kaizu}},
  \bibinfo{author}{\bibfnamefont{B.}~\bibnamefont{Hu}}, \bibnamefont{and}
  \bibinfo{author}{\bibfnamefont{M.}~\bibnamefont{Tomita}},
  \bibinfo{journal}{Bioinformatics} \textbf{\bibinfo{volume}{20}},
  \bibinfo{pages}{538} (\bibinfo{year}{2004}).

\bibitem[{\citenamefont{Vasudeva and Bhalla}(2004)}]{Vasu04}
\bibinfo{author}{\bibfnamefont{K.}~\bibnamefont{Vasudeva}} \bibnamefont{and}
  \bibinfo{author}{\bibfnamefont{U.~S.} \bibnamefont{Bhalla}},
  \bibinfo{journal}{Bioinformatics} \textbf{\bibinfo{volume}{20}},
  \bibinfo{pages}{78} (\bibinfo{year}{2004}).

\bibitem[{\citenamefont{Burrage et~al.}(2004)\citenamefont{Burrage, Tian, and
  Burrage}}]{Burr04}
\bibinfo{author}{\bibfnamefont{K.}~\bibnamefont{Burrage}},
  \bibinfo{author}{\bibfnamefont{T.}~\bibnamefont{Tian}}, \bibnamefont{and}
  \bibinfo{author}{\bibfnamefont{P.}~\bibnamefont{Burrage}},
  \bibinfo{journal}{Prog. Biophys. Mol. Biol.} \textbf{\bibinfo{volume}{85}},
  \bibinfo{pages}{217} (\bibinfo{year}{2004}).

\bibitem[{\citenamefont{Pucha{\l}ka and Kierzek}(2004)}]{Puch04}
\bibinfo{author}{\bibfnamefont{J.}~\bibnamefont{Pucha{\l}ka}} \bibnamefont{and}
  \bibinfo{author}{\bibfnamefont{A.~M.} \bibnamefont{Kierzek}},
  \bibinfo{journal}{Biophys. J.} \textbf{\bibinfo{volume}{86}},
  \bibinfo{pages}{1357} (\bibinfo{year}{2004}).

\bibitem[{\citenamefont{Salis and
  Kaznessis}(2005{\natexlab{a}})}]{Salis05:hybrid}
\bibinfo{author}{\bibfnamefont{H.}~\bibnamefont{Salis}} \bibnamefont{and}
  \bibinfo{author}{\bibfnamefont{Y.}~\bibnamefont{Kaznessis}},
  \bibinfo{journal}{J. Chem. Phys.} \textbf{\bibinfo{volume}{122}},
  \bibinfo{pages}{054103} (\bibinfo{year}{2005}{\natexlab{a}}).

\bibitem[{\citenamefont{Griffith et~al.}(2006)\citenamefont{Griffith, Courtney,
  Peccoud, and Sanders}}]{Griff06}
\bibinfo{author}{\bibfnamefont{M.}~\bibnamefont{Griffith}},
  \bibinfo{author}{\bibfnamefont{T.}~\bibnamefont{Courtney}},
  \bibinfo{author}{\bibfnamefont{J.}~\bibnamefont{Peccoud}}, \bibnamefont{and}
  \bibinfo{author}{\bibfnamefont{W.~H.} \bibnamefont{Sanders}},
  \bibinfo{journal}{Bioinformatics} \textbf{\bibinfo{volume}{22}},
  \bibinfo{pages}{2782} (\bibinfo{year}{2006}).

\bibitem[{\citenamefont{Wylie et~al.}(2006)\citenamefont{Wylie, Hori, Dinner,
  and Chakraborty}}]{Wylie06}
\bibinfo{author}{\bibfnamefont{D.~C.} \bibnamefont{Wylie}},
  \bibinfo{author}{\bibfnamefont{Y.}~\bibnamefont{Hori}},
  \bibinfo{author}{\bibfnamefont{A.~R.} \bibnamefont{Dinner}},
  \bibnamefont{and} \bibinfo{author}{\bibfnamefont{A.~K.}
  \bibnamefont{Chakraborty}}, \bibinfo{journal}{J. Phys. Chem. B}
  \textbf{\bibinfo{volume}{110}}, \bibinfo{pages}{12749}
  (\bibinfo{year}{2006}).

\bibitem[{\citenamefont{Gillespie}(2000)}]{Gillesp00}
\bibinfo{author}{\bibfnamefont{D.~T.} \bibnamefont{Gillespie}},
  \bibinfo{journal}{J. Chem. Phys.} \textbf{\bibinfo{volume}{113}},
  \bibinfo{pages}{297} (\bibinfo{year}{2000}).

\bibitem[{\citenamefont{Chatterjee
  et~al.}(2005{\natexlab{b}})\citenamefont{Chatterjee, Mayawala, Edwards, and
  Vlachos}}]{Chatt05:BioInfo}
\bibinfo{author}{\bibfnamefont{A.}~\bibnamefont{Chatterjee}},
  \bibinfo{author}{\bibfnamefont{K.}~\bibnamefont{Mayawala}},
  \bibinfo{author}{\bibfnamefont{J.~S.} \bibnamefont{Edwards}},
  \bibnamefont{and} \bibinfo{author}{\bibfnamefont{D.~G.}
  \bibnamefont{Vlachos}}, \bibinfo{journal}{Bioinformatics}
  \textbf{\bibinfo{volume}{2005}}, \bibinfo{pages}{2136}
  (\bibinfo{year}{2005}{\natexlab{b}}).

\bibitem[{\citenamefont{Perc et~al.}(2007)\citenamefont{Perc, Gosak, and
  Marhl}}]{Perc07}
\bibinfo{author}{\bibfnamefont{M.}~\bibnamefont{Perc}},
  \bibinfo{author}{\bibfnamefont{M.}~\bibnamefont{Gosak}}, \bibnamefont{and}
  \bibinfo{author}{\bibfnamefont{M.}~\bibnamefont{Marhl}},
  \bibinfo{journal}{Chem. Phys. Lett.} \textbf{\bibinfo{volume}{437}},
  \bibinfo{pages}{143} (\bibinfo{year}{2007}).

\bibitem[{\citenamefont{Handel et~al.}(2007)\citenamefont{Handel,
  {Longini~Jr.}, and Antia}}]{Handel07}
\bibinfo{author}{\bibfnamefont{A.}~\bibnamefont{Handel}},
  \bibinfo{author}{\bibfnamefont{I.~M.} \bibnamefont{{Longini~Jr.}}},
  \bibnamefont{and} \bibinfo{author}{\bibfnamefont{R.}~\bibnamefont{Antia}},
  \bibinfo{journal}{PLoS Comput. Biol.} \textbf{\bibinfo{volume}{3}},
  \bibinfo{pages}{e240} (\bibinfo{year}{2007}).

\bibitem[{\citenamefont{Kummer et~al.}(2000)\citenamefont{Kummer, Olsen, Dixon,
  Green, Bornberg-Bauer, and Baier}}]{Kumm00}
\bibinfo{author}{\bibfnamefont{U.}~\bibnamefont{Kummer}},
  \bibinfo{author}{\bibfnamefont{L.~F.} \bibnamefont{Olsen}},
  \bibinfo{author}{\bibfnamefont{C.~J.} \bibnamefont{Dixon}},
  \bibinfo{author}{\bibfnamefont{A.~K.} \bibnamefont{Green}},
  \bibinfo{author}{\bibfnamefont{E.}~\bibnamefont{Bornberg-Bauer}},
  \bibnamefont{and} \bibinfo{author}{\bibfnamefont{G.}~\bibnamefont{Baier}},
  \bibinfo{journal}{Biophys. J.} \textbf{\bibinfo{volume}{79}},
  \bibinfo{pages}{1188} (\bibinfo{year}{2000}).

\bibitem[{\citenamefont{Berridge et~al.}(1998)\citenamefont{Berridge, Bootman,
  and Lipp}}]{Berr98}
\bibinfo{author}{\bibfnamefont{M.~J.} \bibnamefont{Berridge}},
  \bibinfo{author}{\bibfnamefont{M.~D.} \bibnamefont{Bootman}},
  \bibnamefont{and} \bibinfo{author}{\bibfnamefont{P.}~\bibnamefont{Lipp}},
  \bibinfo{journal}{Nature} \textbf{\bibinfo{volume}{395}},
  \bibinfo{pages}{645} (\bibinfo{year}{1998}).

\bibitem[{\citenamefont{Falcke}(2004)}]{Falcke04}
\bibinfo{author}{\bibfnamefont{M.}~\bibnamefont{Falcke}},
  \bibinfo{journal}{Adv. Phys.} \textbf{\bibinfo{volume}{53}},
  \bibinfo{pages}{255} (\bibinfo{year}{2004}).

\bibitem[{\citenamefont{Schuster et~al.}(2002)\citenamefont{Schuster, Marhl,
  and H\"{o}fer}}]{Schus02}
\bibinfo{author}{\bibfnamefont{S.}~\bibnamefont{Schuster}},
  \bibinfo{author}{\bibfnamefont{M.}~\bibnamefont{Marhl}}, \bibnamefont{and}
  \bibinfo{author}{\bibfnamefont{T.}~\bibnamefont{H\"{o}fer}},
  \bibinfo{journal}{Eur. J. Biochem.} \textbf{\bibinfo{volume}{269}},
  \bibinfo{pages}{1333} (\bibinfo{year}{2002}).

\bibitem[{\citenamefont{Kummer et~al.}(2005)\citenamefont{Kummer, Drajnc,
  Pahle, Green, Dixon, and Marhl}}]{Kumm05}
\bibinfo{author}{\bibfnamefont{U.}~\bibnamefont{Kummer}},
  \bibinfo{author}{\bibfnamefont{B.}~\bibnamefont{Drajnc}},
  \bibinfo{author}{\bibfnamefont{J.}~\bibnamefont{Pahle}},
  \bibinfo{author}{\bibfnamefont{A.~K.} \bibnamefont{Green}},
  \bibinfo{author}{\bibfnamefont{C.~J.} \bibnamefont{Dixon}}, \bibnamefont{and}
  \bibinfo{author}{\bibfnamefont{M.}~\bibnamefont{Marhl}},
  \bibinfo{journal}{Biophys. J.} \textbf{\bibinfo{volume}{89}},
  \bibinfo{pages}{1603} (\bibinfo{year}{2005}).

\bibitem[{\citenamefont{Dixon et~al.}(1990)\citenamefont{Dixon, Woods,
  Cuthbertson, and Cobbold}}]{Dixon90}
\bibinfo{author}{\bibfnamefont{C.~J.} \bibnamefont{Dixon}},
  \bibinfo{author}{\bibfnamefont{N.~M.} \bibnamefont{Woods}},
  \bibinfo{author}{\bibfnamefont{K.~S.~R.} \bibnamefont{Cuthbertson}},
  \bibnamefont{and} \bibinfo{author}{\bibfnamefont{P.~H.}
  \bibnamefont{Cobbold}}, \bibinfo{journal}{Biochem. J.}
  \textbf{\bibinfo{volume}{269}}, \bibinfo{pages}{499} (\bibinfo{year}{1990}).

\bibitem[{\citenamefont{Hasty et~al.}(2001)\citenamefont{Hasty, McMillen,
  Isaacs, and Collins}}]{Hasty01}
\bibinfo{author}{\bibfnamefont{J.}~\bibnamefont{Hasty}},
  \bibinfo{author}{\bibfnamefont{D.}~\bibnamefont{McMillen}},
  \bibinfo{author}{\bibfnamefont{F.}~\bibnamefont{Isaacs}}, \bibnamefont{and}
  \bibinfo{author}{\bibfnamefont{J.~J.} \bibnamefont{Collins}},
  \bibinfo{journal}{Nature Rev. Genet.} \textbf{\bibinfo{volume}{2}},
  \bibinfo{pages}{268} (\bibinfo{year}{2001}).

\bibitem[{\citenamefont{Hasty et~al.}(2002)\citenamefont{Hasty, McMillen, and
  Collins}}]{Hasty02}
\bibinfo{author}{\bibfnamefont{J.}~\bibnamefont{Hasty}},
  \bibinfo{author}{\bibfnamefont{D.}~\bibnamefont{McMillen}}, \bibnamefont{and}
  \bibinfo{author}{\bibfnamefont{J.~J.} \bibnamefont{Collins}},
  \bibinfo{journal}{Nature} \textbf{\bibinfo{volume}{420}},
  \bibinfo{pages}{224} (\bibinfo{year}{2002}).

\bibitem[{\citenamefont{Sprinzak and Elowitz}(2005)}]{Sprin05}
\bibinfo{author}{\bibfnamefont{D.}~\bibnamefont{Sprinzak}} \bibnamefont{and}
  \bibinfo{author}{\bibfnamefont{M.~B.} \bibnamefont{Elowitz}},
  \bibinfo{journal}{Nature} \textbf{\bibinfo{volume}{438}},
  \bibinfo{pages}{443} (\bibinfo{year}{2005}).

\bibitem[{\citenamefont{Benner and Sismour}(2005)}]{Benn05}
\bibinfo{author}{\bibfnamefont{S.~A.} \bibnamefont{Benner}} \bibnamefont{and}
  \bibinfo{author}{\bibfnamefont{A.~M.} \bibnamefont{Sismour}},
  \bibinfo{journal}{Nature Rev. Genet.} \textbf{\bibinfo{volume}{6}},
  \bibinfo{pages}{533} (\bibinfo{year}{2005}).

\bibitem[{\citenamefont{Goldbeter}(2002)}]{Gold02}
\bibinfo{author}{\bibfnamefont{A.}~\bibnamefont{Goldbeter}},
  \bibinfo{journal}{Nature} \textbf{\bibinfo{volume}{420}},
  \bibinfo{pages}{238} (\bibinfo{year}{2002}).

\bibitem[{\citenamefont{Gonze and Goldbeter}(2006)}]{Gonze06}
\bibinfo{author}{\bibfnamefont{D.}~\bibnamefont{Gonze}} \bibnamefont{and}
  \bibinfo{author}{\bibfnamefont{A.}~\bibnamefont{Goldbeter}},
  \bibinfo{journal}{Chaos} \textbf{\bibinfo{volume}{16}},
  \bibinfo{pages}{026110} (\bibinfo{year}{2006}).

\bibitem[{\citenamefont{Sethna}(2006)}]{Sethna06}
\bibinfo{author}{\bibfnamefont{J.~P.} \bibnamefont{Sethna}},
  \emph{\bibinfo{title}{Statistical Mechanics: {E}ntropy, Order Parameters, and
  Complexity}} (\bibinfo{publisher}{Oxford Univ. Press, Oxford, U.K.},
  \bibinfo{year}{2006}), \bibinfo{note}{[{E}xercise (8.11)]}.

\bibitem[{\citenamefont{Sethna and Myers}(2004)}]{Sethna04:hints}
\bibinfo{author}{\bibfnamefont{J.~P.} \bibnamefont{Sethna}} \bibnamefont{and}
  \bibinfo{author}{\bibfnamefont{C.~R.} \bibnamefont{Myers}},
  \emph{\bibinfo{title}{\textit{Entropy, Order Parameters, and Complexity}
  computer exercises: {H}ints and software}} (\bibinfo{year}{2004}),
  \urlprefix\url{http://www.physics.cornell.edu/
  sethna/StatMech/ComputerExercises/Repressilator/Repressilator.html}

\bibitem[{\citenamefont{Cornish-Bowden}(2004)}]{Bowden04}
\bibinfo{author}{\bibfnamefont{A.}~\bibnamefont{Cornish-Bowden}},
  \emph{\bibinfo{title}{Fundamentals of Enzyme Kinetics, 3rd Ed.}}
  (\bibinfo{publisher}{Portland Press Ltd., London, U.K.},
  \bibinfo{year}{2004}).

\bibitem[{\citenamefont{Wallace et~al.}(2004)\citenamefont{Wallace, Kearsley,
  and Guttman}}]{Wall04}
\bibinfo{author}{\bibfnamefont{W.~E.} \bibnamefont{Wallace}},
  \bibinfo{author}{\bibfnamefont{A.~J.} \bibnamefont{Kearsley}},
  \bibnamefont{and} \bibinfo{author}{\bibfnamefont{C.~M.}
  \bibnamefont{Guttman}}, \bibinfo{journal}{Anal. Chem.}
  \textbf{\bibinfo{volume}{76}}, \bibinfo{pages}{2446} (\bibinfo{year}{2004}).

\bibitem[{\citenamefont{Kearsley et~al.}(2005)\citenamefont{Kearsley, Wallace,
  Bernal, and Guttman}}]{Kears05}
\bibinfo{author}{\bibfnamefont{A.~J.} \bibnamefont{Kearsley}},
  \bibinfo{author}{\bibfnamefont{W.~E.} \bibnamefont{Wallace}},
  \bibinfo{author}{\bibfnamefont{J.}~\bibnamefont{Bernal}}, \bibnamefont{and}
  \bibinfo{author}{\bibfnamefont{C.~M.} \bibnamefont{Guttman}},
  \bibinfo{journal}{Appl. Math. Lett.} \textbf{\bibinfo{volume}{18}},
  \bibinfo{pages}{1412} (\bibinfo{year}{2005}).

\bibitem[{not({\natexlab{a}})}]{note:determVar}
\bibinfo{note}{Obviously, deterministic simulations should exhibit zero
  variance in their results. However, due to sampling and curve-fitting
  innaccuracies we do see slight variations. It is these variations that we use
  as the criteria for determining when a system attribute has converged to the
  deterministic limit. Clearly, if the PLA results show equal or less variation
  than the deterministic results then we can deem that the property has
  converged to determinism.}

\bibitem[{\citenamefont{Milton and Arnold}(1995)}]{IntroStats}
\bibinfo{author}{\bibfnamefont{J.~S.} \bibnamefont{Milton}} \bibnamefont{and}
  \bibinfo{author}{\bibfnamefont{J.~C.} \bibnamefont{Arnold}},
  \emph{\bibinfo{title}{Introduction to Probability and Statistics:
  {P}rinciples and Applications for Engineering and the Computing Sciences, 3rd
  Ed.}} (\bibinfo{publisher}{McGraw-Hill Inc., New York, N.Y.},
  \bibinfo{year}{1995}).

\bibitem[{\citenamefont{Cao and Petzold}(2006)}]{Cao06:histDist}
\bibinfo{author}{\bibfnamefont{Y.}~\bibnamefont{Cao}} \bibnamefont{and}
  \bibinfo{author}{\bibfnamefont{L.}~\bibnamefont{Petzold}},
  \bibinfo{journal}{J. Comput. Phys.} \textbf{\bibinfo{volume}{212}},
  \bibinfo{pages}{6} (\bibinfo{year}{2006}).

\bibitem[{not({\natexlab{b}})}]{note:selfDist}
\bibinfo{note}{The self distance is a measure of the difference between a
  sample histogram (i.e., one based on a finite amount of data) and the ``true"
  (unattainable) histogram. Since the measure is based on \emph{absolute\/}
  differences \citep{Cao06:histDist, Harris06} two sample histograms can have
  equal self distances but arising from opposite sources (e.g., one histogram
  might be slightly taller and thinner, while the other shorter and wider, than
  the true histogram). This means that two sample histograms can be as
  dissimilar as \emph{twice\/} the self distance and still be considered
  indistinguishable from the true histogram, and hence each other. In the
  Appendix to Ref.~\citep{Harris06}, it was incorrectly stated that two
  histograms can be considered distinct if they differ by only a single self
  distance.}

\bibitem[{not({\natexlab{c}})}]{note:AdairTimings}
\bibinfo{note}{We found that significant speed-ups can be achieved in the PLA
  simulations of the reduced repressilator model (Table~\ref{table:RepReduced})
  if we removed the ``exact-stochastic" (ES) classification (see
  Fig.~\ref{fig:RepTimings}). The problem lies in the iterative
  $\tau$\/-selection procedure \citep{Harris06} designed to account for the
  randomness of the ES reactions. In this particular case, we experienced an
  unexpected ``classification cascade," whereby reactions classified as ES led
  to a reduced $\tau$\/, which then led to more ES reactions (via
  reclassification), which further reduced $\tau$\/, and so on and so forth.
  Removing the ES classification eliminated this problem with no major effect
  on the accuracy. However, this cannot be done in all cases. Removing the ES
  classification when simulating the full model led to numerous instances of
  negative populations, specifically for the species $g_x$\/, $\{g_x{\cdot}p_r\}$
  and $\{g_x{\cdot}p_r{\cdot}p_r\}$, which can only have populations of zero or unity.
  These required costly reversals that significantly increased the run time.
  Further investigation of this issue is warranted and will be undertaken in
  the near future. Also note that all results reported in
  Figs.~\ref{fig:RepStatsDet} and \ref{fig:RepStatsAdair} were performed
  \emph{with\/} the ES classification included.}

\bibitem[{\citenamefont{Shibata}(2003)}]{Shib03}
\bibinfo{author}{\bibfnamefont{T.}~\bibnamefont{Shibata}}, \bibinfo{journal}{J.
  Chem. Phys.} \textbf{\bibinfo{volume}{119}}, \bibinfo{pages}{6629}
  (\bibinfo{year}{2003}).

\bibitem[{\citenamefont{Bundschuh et~al.}(2003)\citenamefont{Bundschuh, Hayot,
  and Jayaprakash}}]{Bund03}
\bibinfo{author}{\bibfnamefont{R.}~\bibnamefont{Bundschuh}},
  \bibinfo{author}{\bibfnamefont{F.}~\bibnamefont{Hayot}}, \bibnamefont{and}
  \bibinfo{author}{\bibfnamefont{C.}~\bibnamefont{Jayaprakash}},
  \bibinfo{journal}{Biophys. J.} \textbf{\bibinfo{volume}{84}},
  \bibinfo{pages}{1606} (\bibinfo{year}{2003}).

\bibitem[{\citenamefont{Cao et~al.}(2005{\natexlab{b}})\citenamefont{Cao,
  Gillespie, and Petzold}}]{Cao05:slowSSA}
\bibinfo{author}{\bibfnamefont{Y.}~\bibnamefont{Cao}},
  \bibinfo{author}{\bibfnamefont{D.~T.} \bibnamefont{Gillespie}},
  \bibnamefont{and} \bibinfo{author}{\bibfnamefont{L.~R.}
  \bibnamefont{Petzold}}, \bibinfo{journal}{J. Chem. Phys.}
  \textbf{\bibinfo{volume}{122}}, \bibinfo{pages}{014116}
  (\bibinfo{year}{2005}{\natexlab{b}}).

\bibitem[{\citenamefont{Goutsias}(2005)}]{Gout05}
\bibinfo{author}{\bibfnamefont{J.}~\bibnamefont{Goutsias}},
  \bibinfo{journal}{J. Chem. Phys.} \textbf{\bibinfo{volume}{122}},
  \bibinfo{pages}{184102} (\bibinfo{year}{2005}).

\bibitem[{\citenamefont{Samant and Vlachos}(2005)}]{Samant05}
\bibinfo{author}{\bibfnamefont{A.}~\bibnamefont{Samant}} \bibnamefont{and}
  \bibinfo{author}{\bibfnamefont{D.~G.} \bibnamefont{Vlachos}},
  \bibinfo{journal}{J. Chem. Phys.} \textbf{\bibinfo{volume}{123}},
  \bibinfo{pages}{144114} (\bibinfo{year}{2005}).

\bibitem[{\citenamefont{E et~al.}(2005)\citenamefont{E, Liu, and
  Vanden-Eijnden}}]{Weinan05}
\bibinfo{author}{\bibfnamefont{W.}~\bibnamefont{E}},
  \bibinfo{author}{\bibfnamefont{D.}~\bibnamefont{Liu}}, \bibnamefont{and}
  \bibinfo{author}{\bibfnamefont{E.}~\bibnamefont{Vanden-Eijnden}},
  \bibinfo{journal}{J. Chem. Phys.} \textbf{\bibinfo{volume}{123}},
  \bibinfo{pages}{194107} (\bibinfo{year}{2005}).

\bibitem[{\citenamefont{Salis and
  Kaznessis}(2005{\natexlab{b}})}]{Salis05:QSSA}
\bibinfo{author}{\bibfnamefont{H.}~\bibnamefont{Salis}} \bibnamefont{and}
  \bibinfo{author}{\bibfnamefont{Y.~N.} \bibnamefont{Kaznessis}},
  \bibinfo{journal}{J. Chem. Phys.} \textbf{\bibinfo{volume}{123}},
  \bibinfo{pages}{214106} (\bibinfo{year}{2005}{\natexlab{b}}).

\bibitem[{\citenamefont{Morelli et~al.}(2008)\citenamefont{Morelli, Allen,
  {T\u{a}nase-Nicola}, and {ten~Wolde}}}]{Morelli08}
\bibinfo{author}{\bibfnamefont{M.~J.} \bibnamefont{Morelli}},
  \bibinfo{author}{\bibfnamefont{R.~J.} \bibnamefont{Allen}},
  \bibinfo{author}{\bibfnamefont{S.}~\bibnamefont{{T\u{a}nase-Nicola}}},
  \bibnamefont{and} \bibinfo{author}{\bibfnamefont{P.~R.}
  \bibnamefont{{ten~Wolde}}}, \bibinfo{journal}{J. Chem. Phys.}
  \textbf{\bibinfo{volume}{128}}, \bibinfo{pages}{045105}
  (\bibinfo{year}{2008}).

\bibitem[{\citenamefont{Hlavacek et~al.}(2006)\citenamefont{Hlavacek, Faeder,
  Blinov, Posner, Hucka, and Fontana}}]{Hlava06}
\bibinfo{author}{\bibfnamefont{W.~S.} \bibnamefont{Hlavacek}},
  \bibinfo{author}{\bibfnamefont{J.~R.} \bibnamefont{Faeder}},
  \bibinfo{author}{\bibfnamefont{M.~L.} \bibnamefont{Blinov}},
  \bibinfo{author}{\bibfnamefont{R.~G.} \bibnamefont{Posner}},
  \bibinfo{author}{\bibfnamefont{M.}~\bibnamefont{Hucka}}, \bibnamefont{and}
  \bibinfo{author}{\bibfnamefont{W.}~\bibnamefont{Fontana}},
  \bibinfo{journal}{Sci. STKE} \textbf{\bibinfo{volume}{2006
  \textnormal{(344)}}}, \bibinfo{pages}{re6} (\bibinfo{year}{2006}).

\bibitem[{\citenamefont{Lemerle et~al.}(2005)\citenamefont{Lemerle,
  {Di~Ventura}, and Serrano}}]{Lemerle05}
\bibinfo{author}{\bibfnamefont{C.}~\bibnamefont{Lemerle}},
  \bibinfo{author}{\bibfnamefont{B.}~\bibnamefont{{Di~Ventura}}},
  \bibnamefont{and} \bibinfo{author}{\bibfnamefont{L.}~\bibnamefont{Serrano}},
  \bibinfo{journal}{FEBS Lett.} \textbf{\bibinfo{volume}{579}},
  \bibinfo{pages}{1789} (\bibinfo{year}{2005}).

\bibitem[{\citenamefont{Brown and Sethna}(2003)}]{Brown03}
\bibinfo{author}{\bibfnamefont{K.~S.} \bibnamefont{Brown}} \bibnamefont{and}
  \bibinfo{author}{\bibfnamefont{J.~P.} \bibnamefont{Sethna}},
  \bibinfo{journal}{Phys. Rev. E} \textbf{\bibinfo{volume}{68}},
  \bibinfo{pages}{021904} (\bibinfo{year}{2003}).

\bibitem[{\citenamefont{Gunawan et~al.}(2005)\citenamefont{Gunawan, Cao,
  Petzold, and {Doyle~III}}}]{Guna05}
\bibinfo{author}{\bibfnamefont{R.}~\bibnamefont{Gunawan}},
  \bibinfo{author}{\bibfnamefont{Y.}~\bibnamefont{Cao}},
  \bibinfo{author}{\bibfnamefont{L.}~\bibnamefont{Petzold}}, \bibnamefont{and}
  \bibinfo{author}{\bibfnamefont{F.~J.} \bibnamefont{{Doyle~III}}},
  \bibinfo{journal}{Biophys.~J.} \textbf{\bibinfo{volume}{88}},
  \bibinfo{pages}{2530} (\bibinfo{year}{2005}).

\bibitem[{\citenamefont{Palsson}(2000)}]{Palsson00}
\bibinfo{author}{\bibfnamefont{B.}~\bibnamefont{Palsson}},
  \bibinfo{journal}{Nat. Biotechnol.} \textbf{\bibinfo{volume}{18}},
  \bibinfo{pages}{1147} (\bibinfo{year}{2000}).

\end{thebibliography}

\end{document}